\documentclass[structabstract]{aa}

\usepackage{textcomp}

\usepackage{graphicx}
\usepackage{txfonts}
\usepackage{natbib}
\bibpunct{(}{)}{;}{a}{}{,} 

\newcommand\nobrkhyph{\mbox{-}}

\begin{document}
   \title{A Survey of HC$_3$N in Extragalactic Sources}

   \subtitle{Is HC$_3$N a Tracer of Activity in ULIRGs?}

   \author{J. E. Lindberg
          \inst{1,2,3}\thanks{J.L. wishes to thank Instrumentcenter for Danish Astrophysics (IDA) for grant support.}
          \and
          S. Aalto\inst{3}\thanks{S.A. wishes to thank the Swedish Research Council for grant support.}
          \and
          F. Costagliola\inst{3}\thanks{F.C. wishes to thank the EU ESTRELA programme for support.}
          \and
          J.-P. P\'{e}rez-Beaupuits\inst{4,5}
          \and
          R. Monje\inst{6}
          \and
          S. Muller\inst{3}
          }

   \institute{{Centre for Star and Planet Formation, Natural History Museum of Denmark, University of Copenhagen, {\O}ster Voldgade 5-7, DK\nobrkhyph1350 Copenhagen K, Denmark}
   			\email{jlindberg@snm.ku.dk}
         \and
         	{Nordic Optical Telescope, Apartado 474, E-38700 Santa Cruz de La Palma, Santa Cruz de Tenerife, Spain}
         \and
         	{Department of Earth and Space Sciences, Onsala Observatory, Chalmers University of Technology, SE-439 92 Onsala, Sweden}
         \and
            {Max-Planck-Institut f\"ur Radioastronomie, Auf dem H\"ugel 69, 53121 Bonn, Germany}
         \and
            {Kapteyn Astronomical Institute, University of Groningen, Landleven 12, 9747 AD Groningen, The Netherlands}
         \and
         	{California Institute of Technology, 1200 E. California Blvd., Mail Code 301-17, Pasadena, CA  91125-4700, USA}\\
        }
     
   \date{Received August 11, 2010; accepted December 16, 2010}

 
  \abstract
   {HC$_3$N is a molecule that is mainly associated with Galactic star-forming regions, but it has also been detected in extragalactic environments.} 	
   {To present the first extragalactic survey of HC$_3$N, when combining earlier data from the literature with six new single-dish detections, and to compare HC$_3$N with other molecular tracers (HCN, HNC), as well as other properties (silicate absorption strength, IR flux density ratios, \ion{C}{ii} flux, and megamaser activity).}
   {We present mm IRAM~30~m, OSO~20~m, and SEST observations of HC$_3$N rotational lines (mainly the $J=$10\nobrkhyph9 transition) and of the $J=$1\nobrkhyph0 transitions of HCN and HNC. Our combined HC$_3$N data account for 13 galaxies (excluding the upper limits reported for the non-detections), while we have HCN and HNC data for more than 20 galaxies.}
   {A preliminary definition ``HC$_3$N-luminous galaxy'' is made based upon the HC$_3$N/HCN ratio. Most ($\sim$80~$\%$) HC$_3$N-luminous galaxies seem to be deeply obscured galaxies and (U)LIRGs. A majority ($\sim$60~$\%$ or more) of the HC$_3$N-luminous galaxies in the sample present OH mega- or strong kilomaser activity. A possible explanation is that both HC$_3$N and OH megamasers need warm dust for their excitation. Alternatively, the dust that excites the OH megamaser offers protection against UV destruction of HC$_3$N. A high silicate absorption strength is also found in several of the HC$_3$N-luminous objects, which may help the HC$_3$N to survive. Finally, we find that a high HC$_3$N/HCN ratio is related to a high dust temperature and a low \ion{C}{ii} flux.}
   {}


   \maketitle
%

\section{Introduction}

Finding useful tracers of the interaction between the activity in galaxy nuclei and surrounding interstellar medium (ISM) is
an important and growing aspect of current extragalactic molecular astronomy. In this context, single dish surveys of polar molecules
such as HCN, HNC, HCO$^+$, and CS have been used to investigate possible correlations between molecular line ratios and type/intensity of activity \citep[e.g.][]{kohno01,aalto02,imanishi04,gracia06,krips08,baan08}.
For example, it has been suggested that an elevated HCN/HCO$^+$~1\nobrkhyph0 line intensity ratio indicates the presence of an AGN \citep{gracia06}. 
Around an active galactic nucleus (AGN) the chemistry is supposedly dominated by hard X-rays in an X-ray dominated region (XDR), and some chemical
models predict an abundance enhancement of HCN paired with selective destruction of HCO$^+$ \citep{maloney96} -- which could lead to an
elevated HCN/HCO$^+$ line ratio (under the circumstances that the line ratio directly reflects the abundance ratio).
However, more recent chemical models instead suggest that HCO$^+$ is enhanced in XDRs \citep{meijerink05,meijerink07}, and HCO$^+$ is also expected to be under-abundant in
regions of very young star formation \citep{aalto08}, so the line ratio is ambiguous.

Other molecular tracers could help resolve the dichotomy of the HCN/HCO$^+$ line ratio. The serendipitous discovery of the $J$=10\nobrkhyph9 line
of HC$_3$N near the HNC~$J$=1\nobrkhyph0 line in a survey by \citet{aalto02} led us to look more closely at this molecule. HC$_3$N is the
simplest of the cyanopolyynes (carbon chains with an attached CN group) and is a grain chemistry product, in contrast to
molecules such as HCN and HCO$^+$. HC$_3$N thrives in warm, dense shielded regions such as hot cores where abundances
can reach 10$^{-8}$ or even higher, since it is easily destroyed by photo-dissociation \citep{rodriguez98} and C$^+$ ions \citep{prasad80}.
Therefore, HC$_3$N line emission could be used to identify galaxies where star formation 
is in the early, embedded stage of its evolution. Recently, HC$_3$N was found in high abundance in the highly obscured galaxy \object{NGC 4418}
\citep{aalto07a}, as well as the ULIRG \object{Arp 220} \citep{aalto02}.

We have searched for HC$_3$N line emission in a sample of galaxies in various stages and types of activity:
AGNs, starbursts, and ultraluminous galaxies (ULIRGs). In some of the galaxies the nature of the activity is elusive
since it is embedded in huge columns of dust absorbing emission at optical and infrared wavelengths. In some cases, the extinction
is so strong that no emission emerges at optical or IR wavelengths requiring us to probe the nature of the activity at radio
and mm wavelengths. HC$_3$N has a rich mm and sub-mm wavelength spectrum
consisting of a multitude of rotational and vibrational lines
often appearing close to each other in the same band.
Through its vibrational transitions, HC$_3$N responds strongly
to the IR field from dusty nuclei \citep{costagliola10a}. Therefore, combining the rotational and vibrational line information
of HC$_3$N allows us to study the abundance of HC$_3$N
(comparing with chemical models of XDRs and starbursts) as well as
the intensity and temperature structure of the buried IR source.

Rotational lines of vibrationally excited HC$_3$N have recently been discovered in a few galaxies (\object{NGC 4418} \citep{costagliola10a}, \object{Arp 220} \citep{martin10}, and \object{IC 860} \citep{costagliola10b}), therefore showing that it is important to take both radiative and collisional excitation into consideration when interpreting HC$_3$N line emission from IR luminous galaxies.

It can also be noted that absorption lines of HC$_3$N has been found in a $z\sim0.89$ galaxy located in front of the quasar \object{PKS 1830-211} \citep{henkel09}.

\subsection{Outline}

Here, the first survey of extragalactic HC$_3$N data is presented. We report new HC$_3$N observations in 19 galaxies (detections in six of them), mainly (U)LIRGs and starburst galaxies, and complete this sample with data from all earlier extragalactic HC$_3$N emission line single-dish detections found in the literature. The aim of the study is to compare the HC$_3$N luminosity with other molecular tracers as well as galaxy properties to see if the presence of HC$_3$N can be used to predict other galaxy properties, e.g. the source of activity in the galaxy.

In Section~\ref{sec:hc3ninspace}, the general properties of HC$_3$N in space are discussed. In Section~\ref{sec:obs} we present the new observations and discuss the collection of data from the literature. In Section~\ref{sec:res} we present the results in terms of line intensities and line ratios. In Section~\ref{sec:dis} we discuss the interpretation of the HC$_3$N results and compare them with silicate absorption strength (Section~\ref{sec:pah}), OH megamaser activity (Section~\ref{sec:meg}), IR flux density ratios (Section~\ref{sec:iras}), \ion{C}{ii} flux (Section~\ref{sec:cii}), and the HNC/HCN~1\nobrkhyph0 line ratio (Section~\ref{sec:hnc}). In Section~\ref{sec:fut} future studies resulting from this project are discussed.

\section{HC$_3$N in space}
\label{sec:hc3ninspace}

\subsection{Generation of HC$_3$N}
\label{sec:c2h2c2h}

Acetylene, C$_2$H$_2$, exists on grains in the ISM \citep{chapman09}. After evaporating from the grains there are at least two different paths the C$_2$H$_2$ may follow. If a high UV field is present (the region being a PDR), it will photo-dissociate into the ethynyl radical, C$_2$H \citep{meier05,cherchneff93,heikkila99}:

\begin{equation}
\indent \mathrm{C}_2\mathrm{H}_2 + h \nu \longrightarrow \mathrm{C}_2\mathrm{H} + \mathrm{H}.
\end{equation}

\noindent
If no strong UV field is present (no PDR), but CN (the cyano radical) is available for reactions, the C$_2$H$_2$ will instead react with the CN to form HC$_3$N \citep{meier05,chapman09,fukuzawa97}:

\begin{equation}
\indent \mathrm{C}_2\mathrm{H}_2 + \mathrm{CN} \longrightarrow \mathrm{HC}_3\mathrm{N} + \mathrm{H}.
\end{equation}

\noindent
This hypothesis is strengthened by interferometric maps of HC$_3$N and C$_2$H in \object{IC 342} found in \citet{meier05}. The maps show a clear anti-correlation between the distributions of the molecules. Using the abundances of HC$_3$N, C$_2$H, and CN in a region where it is expected that grains with C$_2$H$_2$ first were present, it should therefore be possible to tell whether a strong UV field is present or not.

\citet{irvine87} find that C$_2$H, contrarily to HC$_3$N, is two orders of magnitude more abundant in the Orion ridge than in its hot core. Possibly, the hot core shields the HC$_3$N and C$_2$H$_2$ from photo-dissociating into C$_2$H. Some of the C$_2$H$_2$ instead reacts with the CN (although not very abundant) and forms even more HC$_3$N. The highest HC$_3$N abundances are found in \object{Sgr B2} hot cores, being in the order of $10^{-7}$ of the H$_2$ abundance \citep{devicente00}.

We have not included all possible HC$_3$N formation mechanisms here, and investigations of other processes are ongoing, as for example the notion of ice formation of HC$_3$N \citep{szczepanski05}.

\subsection{Destruction of HC$_3$N}
\label{sec:destruct}

In the Galaxy, HC$_3$N is associated with warm, dense, shielded gas around young stars or star-forming regions, and is easily destroyed by UV radiation and reactions with C$^+$ ions \citep{rodriguez98,meier05}. It will form either C$_2$H or C$_3$N when being photo-dissociated \citep{cherchneff93}, and C$_3$H$^+$ or C$_4$N$^+$ when reacting with C$^+$ \citep{bohme85}. Among the possible reactions destroying HC$_3$N are:

\begin{equation}
\indent \mathrm{HC}_3\mathrm{N} + h \nu \longrightarrow \mathrm{C}_2\mathrm{H} + \mathrm{CN},
\end{equation}

\begin{equation}
\indent \mathrm{HC}_3\mathrm{N} + h \nu \longrightarrow \mathrm{C}_3\mathrm{N} + \mathrm{H},
\end{equation}

\begin{equation}
\label{eq:cii1}
\indent \mathrm{HC}_3\mathrm{N} + \mathrm{C}^+ \longrightarrow \mathrm{C}_3\mathrm{H}^+ + \mathrm{CN},
\end{equation}

\begin{equation}
\label{eq:cii2}
\indent \mathrm{HC}_3\mathrm{N} + \mathrm{C}^+ \longrightarrow \mathrm{C}_4\mathrm{N}^+ + \mathrm{H}.
\end{equation}

\noindent
Reaction rates of these reactions, as well as those in Section~\ref{sec:c2h2c2h}, can be found in e.g. \citet{cherchneff93}.

\subsection{Abundances of HC$_3$N}

\citet{irvine87} give the relative abundances of several molecules in the core and ridge of the Orion molecular cloud. The detections of these molecules in the Galaxy are indicative of their abundances in high and low density molecular regions. The relative HC$_3$N abundance lies around $10^{-9}$ of the H$_2$ abundance in the core (dense region), and $10^{-10}$ in the ridge (low density region). This is a relatively small difference between high and low density regions, as compared to e.g. HCN, with about $10^{-7}$ of the H$_2$ abundance in the core, and $10^{-9}$ in the ridge.

The intense radiation from starburst regions and/or AGNs in the centre of many galaxies will turn surrounding gas clouds into regions where the chemical structure depends highly on the radiation field, either photon-dominated regions (PDRs) or X\nobrkhyph ray dominated regions (XDRs). In XDRs the abundance of several molecules (e.g. CN and CH$_2$) are expected to be enhanced with respect to the abundance commonly found in PDRs. Due to the high C$^+$ abundance in XDRs, a low HC$_3$N abundance is expected \citep{aalto08}. The HC$_3$N/CN abundance ratio found in PDRs is also very low compared to the same ratio measured in hot, dense cores \citep{rodriguez98}. For further discussion about PDR and XDR chemistry, see e.g. \citet{tielens85,lepp96,meijerink05,maloney96}.


\section{Observations}
\label{sec:obs}

The new observations reported in this work were carried out with the IRAM~30~m\footnote{Based on observations carried out with the IRAM~30m Telescope. IRAM is supported by INSU/CNRS (France), MPG (Germany), and IGN (Spain)}, OSO~20~m, and SEST~15~m telescopes between 2001 and 2008. Detailed lists with system temperatures and dates for the observations reported in this work are found in Table~\ref{tab:obsdates}). The pointing accuracy was better than $2''$ for all the observations, and typical system temperatures were 150~K (IRAM 90\nobrkhyph110~GHz), 250~K (IRAM 125~GHz), 400~K (IRAM 225\nobrkhyph250~GHz), and 300~K (OSO and SEST). In this survey we also include data from the literature using the already mentioned telescopes, as well as the NRO~45~m, NRAO~12~m, and FCRAO~14~m telescopes. Whenever using data from the literature, the beam sizes and efficiencies given in the respective articles have been used for calculations of line ratios. The parameters used for all new observations reported in this work are given in Table~\ref{tab:obsparam}.

One could argue that the many different instruments used to obtain the data in this article might introduce a bias difficult to compensate for. However, when comparing the HNC/HCN~1\nobrkhyph0 line ratios obtained with IRAM and SEST respectively, no systematic bias is detected. The average line ratio was calculated to 0.51$\pm$0.11 for SEST data and 0.51$\pm$0.08 for IRAM data. 

\begin{table}
\centering
      \caption[]{Data for the observations reported in this work}
         \label{tab:obsdates}
         \begin{tabular}{l l l l l}
 
            \noalign{\smallskip}
			\hline
            \hline
            \noalign{\smallskip}
            Galaxy & Molecule & Line & Telescope & Date \\
            \noalign{\smallskip}
            \hline
            \noalign{\smallskip}
			\object{Arp 220} & HC$_3$N & 10-9 & OSO & 2001-11-11 \\
			\object{Arp 220} & HC$_3$N & 12-11 & OSO & 2001-11-09 \\
			\object{Circinus} & HNC & 1-0 & SEST & 2001-01-14 \\
			\object{IC 694} & HC$_3$N & 12-11 & OSO & 2001-11-10 \\
			\object{IC 860} & HC$_3$N & 28-27 & IRAM & 2007-12-16 \\
			\object{IC 860} & HNC & 1-0 & IRAM & 2007-12-16 \\
			I17208 & HC$_3$N & 10-9 & IRAM & 2007-12-16 \\
			I17208 & HNC & 1-0 & IRAM & 2006-06-28 \\
			\object{Maffei 2} & HC$_3$N & 12-11 & IRAM & 2007-08-25 \\
			\object{NGC 34} & HNC & 1-0 & SEST & 2001-01-14 \\
			\object{NGC 613} & HCN & 1-0 & SEST & 2001-02-11 \\
			\object{NGC 613} & HNC & 1-0 & SEST & 2001-02-13 \\
			\object{NGC 1056} & HC$_3$N & 16-15 & IRAM & 2006-06-30 \\
			\object{NGC 1056} & HNC & 1-0 & IRAM & 2006-06-30 \\
			\object{NGC 1377} & HC$_3$N & 16-15 & IRAM & 2007-12-13 \\
			\object{NGC 1377} & HC$_3$N & 25-24 & IRAM & 2007-12-13 \\
			\object{NGC 1377} & HCN & 1-0 & IRAM & 2007-12-13 \\
			\object{NGC 1377} & HNC & 1-0 & IRAM & 2007-12-13 \\
			\object{NGC 1614} & HNC & 1-0 & SEST & 2001-02-13 \\
			\object{NGC 2146} & HC$_3$N & 10-9 & OSO & 2001-11-08 \\
			\object{NGC 2146} & HC$_3$N & 12-11 & OSO & 2001-11-08 \\
			\object{NGC 2623} & HC$_3$N & 12-11 & OSO & 2001-11-12 \\
			\object{NGC 3079} & HC$_3$N & 10-9 & IRAM & 2006-05-14 \\
			\object{NGC 3079} & HC$_3$N & 16-15 & IRAM & 2006-05-14 \\
			\object{NGC 3079} & HC$_3$N & 25-24 & IRAM & 2006-05-14 \\
			\object{NGC 3690} & HC$_3$N & 12-11 & OSO & 2001-11-09 \\
			\object{NGC 4418} & HCN & 1-0 & IRAM & 2008-07-19 \\
			\object{NGC 4945} & HNC & 1-0 & SEST & 2001-01-15 \\
			\object{NGC 5135} & HNC & 1-0 & SEST & 2001-01-13 \\
			\object{NGC 6946} & HC$_3$N & 12-11 & IRAM & 2007-08-25 \\
			\object{UGC 5101} & HC$_3$N & 10-9 & IRAM & 2007-12-13 \\
			\object{UGC 5101} & HNC & 1-0 & IRAM & 2007-12-13 \\
			
            \noalign{\smallskip}
            \hline

     \end{tabular}
     \tablefoot{
     	 For the observations cited from the literature, see the respective articles referred to in Tables~\ref{tab:resultshc3n}\nobrkhyph\ref{tab:resultshnc}. We would like to point out that some of the HC$_3$N~10\nobrkhyph9 data reported by us come from SEST HNC~1\nobrkhyph0 spectra.
	 }
\end{table}

\begin{table}
\centering
      \caption[]{Observational parameters.}
         \label{tab:obsparam}
         \begin{tabular}{l r r r}
 
            \noalign{\smallskip}
			\hline
            \hline
            \noalign{\smallskip}
            Transition & $\nu$ [GHz]\tablefootmark{a} & HPBW [$''$]\tablefootmark{b} & ${\eta_{\mathrm{mb}}}$\tablefootmark{b} \\
            \noalign{\smallskip}
            \hline
            \noalign{\smallskip}
            \textit{IRAM:} \\
            HC$_3$N~10\nobrkhyph9 & 90.979 & 28 & 0.80 \\ 
            HC$_3$N~12\nobrkhyph11 & 109.174 & 24 & 0.73 \\
            HC$_3$N~16\nobrkhyph15 & 145.561 & 17 & 0.67 \\
            HC$_3$N~25\nobrkhyph24 & 227.419 & 10.5 & 0.63 \\
            HC$_3$N~28\nobrkhyph27 & 254.699 & 9 & 0.59 \\
            HCN~1\nobrkhyph0 & 88.632 & 28 & 0.80 \\
            HNC~1\nobrkhyph0 & 90.664 & 28 & 0.80 \\
            
            \textit{OSO:} \\
            HC$_3$N~12\nobrkhyph11 & 109.174 & 36 & 0.52 \\
            HC$_3$N~10\nobrkhyph9 & 90.979 & 42 & 0.59 \\
            
            \textit{SEST:} \\
            HC$_3$N~10\nobrkhyph9 & 90.979 & 55 & 0.75 \\
            HCN~1\nobrkhyph0 & 88.632 & 57 & 0.75 \\
            HNC~1\nobrkhyph0 & 90.664 & 55 & 0.75 \\
            
            \noalign{\smallskip}
            \hline

     	\end{tabular}
     	\tablefoot{
     	\tablefoottext{a}{From the NIST database \textit{Recommended Rest Frequencies for Observed Interstellar Molecular Microwave Transitions} (http://physics.nist.gov/cgi-bin/micro/table5/start.pl).} \\
     	\tablefoottext{b}{The half-power beamwidths and main beam efficiencies are collected from the respective telescope web pages.}
     	}
\end{table}

All objects investigated (observed by us or with data from the literature) are listed in Table~\ref{tab:galaxies}, along with some important characteristics.

The relative HC$_3$N abundances calculated in this work will be expressed as line ratios between an HC$_3$N line (mostly the $J=$10\nobrkhyph9 transition) and the $J=$1\nobrkhyph0 transitions of HCN and HNC. These two molecules are chosen as they are good tracers of high density regions \citep[see e.g.][]{papadopoulos07,aalto02}, where we expect to find the HC$_3$N \citep{meier05}. Also, HCN and HNC data are available for most of the objects in the sample. We note that the line ratios are not linearly proportional to ratios between the abundances of the species, since they will also depend on excitation conditions and optical depths in the galaxies. A high HC$_3$N/HCN ratio might thus sometimes be a tracer of discrepancies in temperatures, densities, or IR pumping in the galaxies.

A discussion of the method used when calculating the line ratios can be found in Appendix~\ref{app:nearby}. A few of the most nearby galaxies in the survey have such a large angular distribution that the measured values might not represent a global value for molecular gas in the galaxy, but rather a value for a certain (central) region of the galaxy. This effect is discussed in Appendix~\ref{app:nearby2}.

The sample of galaxies observed by us has been chosen to have a high probability of finding HC$_3$N -- it is by no means intended to be an unbiased sample of some random galaxies, and thus the relatively high detection ratio should definitely not reflect the amount of HC$_3$N-luminous galaxies in the universe. The same is most likely true for the galaxies found in the literature. Another important selection effect for the objects from the literature is that detections are much more likely to be reported than non-detections, which also leads to a biased sample.

To increase the chance of detecting HC$_3$N, almost all of the galaxies that were chosen to be part of the sample have earlier detections of HCN, which means that they should have large amounts of dense gas, increasing the possibility of finding HC$_3$N. As the goal of the study is to investigate if HC$_3$N can trace the source of the activity in active galaxies, the sample consists only of active galaxies -- starburst galaxies and AGN galaxies (the source of the activity is although disputed or unknown in many of the galaxies in the sample).

Throughout the article, the $T_{\mathrm{A}}^*$ scale will be used for all our data. For the data from the literature, the temperature scale used in each article will be used in our tables, clearly noted whenever the $T_{\mathrm{A}}^*$ scale is not used. This will make it easier to detect any errors that might have occurred in the survey work. When the line ratios are calculated, the efficiencies will be taken into account properly.

Data analysis was performed with the X-Spec\footnote{http://www.chalmers.se/rss/oso-en/observations/data-reduction-software} software package. A first order baseline was subtracted from all spectra.

\begin{table*}
\centering
      \caption[]{List of investigated objects and some of their properties. For the galaxies with new observations reported, the given positions are those used for our observations. For objects not observed by us the position in \citet{ned} is given.}
         \label{tab:galaxies}
         \begin{tabular}{l l l l r r r r}
 
            \noalign{\smallskip}
			\hline
            \hline
            \noalign{\smallskip}
            Galaxy & R.A. & Decl. & Type\tablefootmark{a} & $cz$\tablefootmark{b} & $D$\tablefootmark{c} & ${\log L_{\mathrm{FIR}}}$\tablefootmark{d} & ${\theta_{\mathrm{HCN}}}$\tablefootmark{e} \\
            & (J2000.0) & (J2000.0) & & [km s$^{-1}$] & [Mpc] & [$L_{\odot}$] & [$''$] \\
            \noalign{\smallskip}
            \hline
            \noalign{\smallskip}
			\object{Arp 220} & 15 34 57.1 & +23 30 11.3 & ULIRG, Obsc., SB? & 5450 & 78.1 & 12.15 & 2$^1$ \\
			\object{Circinus} & 14 10 34.3 & --64 52 12.5 & AGN, cp? & 434\tablefootmark{f} & 3.13 & ... & 21 (CO~3\nobrkhyph2)$^2$ \\
			\object{IC 342} & 03 46 48.5 & +68 05 46 & SB & 31 & 4.00 & 10.01 & 20$^3$ \\
			\object{IC 694}\tablefootmark{g} & 11 28 33.6 & +58 33 46.0 & SB & 3159 & 58.2 & 11.74\tablefootmark{h} & 5$^4$ \\
			\object{IC 860} & 13 15 03.5 & +24 37 08.0 & Obsc. & 3887 & 59.1\tablefootmark{i} & 11.14 & * \\
			I17208\tablefootmark{j} & 17 23 21.9 & --00 17 00.9 & ULIRG, Obsc., SB? & 12852 & 178 & 12.35 & 0.67$^5$ \\
			\object{M82} & 09 55 52.7 & +69 40 46 & SB & 187 & 5.68 & 10.61 & $>30^1$ \\
			\object{Maffei 2} & 02 41 55.1 & +59 36 15.0 & SB & -17\tablefootmark{f} & 3.34 & ... & $20 \times 7^6$ \\
			\object{NGC 34} & 00 11 06.5 & --12 06 26.6 & SB & 5931 & 79.8 & 11.34 & * \\
			\object{NGC 253} & 00 47 33.1 & --25 17 18 & SB & 261 & 3.22 & 10.29 & $18 \times 8^7$ \\
			\object{NGC 613} & 01 36 36.7 & --29 09 50.4 & cp & 1475 & 18.6 & 10.22 & 40 (CO~1\nobrkhyph0)$^8$ \\
			\object{NGC 1056} & 02 42 48.3 & +28 34 27.1 & AGN & 1545 & 22.7 & 9.79 & - \\
			\object{NGC 1068} & 02 42 40.7 & --00 00 47.0 & cp & 1005 & 15.3 & 10.89 & 4.5$^1$\\
			\object{NGC 1365} & 03 33 36.4 & --36 08 26.1 & cp & 1636 & 19.9 & 10.86 & 34 (CO~2\nobrkhyph1)$^9$ \\
			\object{NGC 1377} & 03 36 39.1 & --20 54 08.0 & Obsc., AGN? & 1792 & 22.5 & 9.95 & - \\
			\object{NGC 1614} & 04 36 24.2 & --08 28 40.3 & SB & 4746 & 63.4 & 11.43 & 12 (CO~2\nobrkhyph1)$^{10}$ \\
			\object{NGC 1808} & 05 07 42.3 & --37 30 47 & SB, cp? & 1000 & 11.2 & 10.55 & 18 (CO~2\nobrkhyph1)$^{10}$ \\
			\object{NGC 2146} & 06 18 37.8 & +78 21 22.9 & SB & 885 & 16.9 & 10.93 & 20$^1$ \\
			\object{NGC 2623} & 08 38 24.1 & +25 45 17.2 & SB & 5538 & 76.9 & 11.48 & 1.8$^{11}$ \\
			\object{NGC 3079} & 10 01 57.81 & +55 40 47.1 & SB? AGN? & 1142 & 19.7 & 10.65 & $13 \times 5$ (CO~1\nobrkhyph0)$^{12}$ \\
			\object{NGC 3256} & 10 27 51.3 & --43 54 14 & SB & 2781 & 36.5 & 11.43 & 9 (CO~2\nobrkhyph1)$^{10}$ \\
			\object{NGC 3690}\tablefootmark{g} & 11 28 31.0 & +58 33 40.0 & SB & 3159 & 46.9 & 11.74\tablefootmark{h} & 1.56$^5$ \\
			\object{NGC 4418} & 12 26 54.8\tablefootmark{k} & --00 52 42.0\tablefootmark{k} & Obsc., AGN? & 2104 & 32.6 & 11.00 & 5 (CO~1\nobrkhyph0)$^{13}$\\
			\object{NGC 4945} & 13 05 27.0 & --49 28 04.5 & SB, cp? & 560 & 4.85 & 10.41 & 15 (CO~3\nobrkhyph2)$^2$ \\
			\object{NGC 5135} & 13 25 44.0 & --29 50 02.2 & cp & 4114 & 56.0 & 11.06 & $15 \times 5$ (CO~1\nobrkhyph0)$^{14}$\\
			\object{NGC 6946} & 20 34 52.3 & +60 09 14.0 & SB & 53 & 5.64 & 10.01 & 10$^1$ \\
			\object{NGC 7130} & 21 48 19.5 & --34 57 05 & cp & 4824 & 65.4 & 11.23 & 10 (CO~1\nobrkhyph0)$^{10}$ \\
			\object{UGC 5101} & 09 35 51.6 & +61 21 11.7 & LIRG, cp & 11785 & 165 & 11.87 & 3.50 (CO~1\nobrkhyph0)$^5$ \\
            \noalign{\smallskip}
            \hline

     \end{tabular}
     
     \tablebib{(1) Table in \citet{krips08}; (2) Table in \citet{curran01a}; (3) HCN map in \citet{meier05}; (4) HCN map in \citet{aalto97}; (5) Table in \citet{gracia08}; (6) HCN map in \citet{nguyen94}; (7) HCN map in \citet{knudsen07}; (8) CO~1\nobrkhyph0 map in \citet{bajaja95}; (9) CO~2\nobrkhyph1 source size in \citet{curran01b}; (10) Table in \citet{aalto95}; (11) Table in \citet{bryant99}; (12) CO~1\nobrkhyph0 map in \citet{koda02}; (13) CO~1\nobrkhyph0 map in \citet{dale05}; (14) CO~1\nobrkhyph0 map in \citet{regan99}.
      }
 	 \tablefoot{
 	 \tablefoottext{a}{The classifications have been obtained by careful investigation of the notes in \citet{ned}. SB = starburst, AGN = active galactic nucleus, cp  = composite of SB and AGN, Obsc. = obscured, ULIRG = Ultra-luminous Infrared galaxy, LIRG = Luminous Infrared galaxy.}
 	 \tablefoottext{b}{Heliocentric radial velocity of source, from \citet{sanders03}.}
 	 \tablefoottext{c}{Distance to source, corrected for Virgo infall only, from \citet{ned}.}
 	 \tablefoottext{d}{Far Infrared Luminosity, from \citet{sanders03}.}
 	 \tablefoottext{e}{Source sizes, given for HCN~1\nobrkhyph0 line if not specified otherwise. For galaxies with an asterisk (*), no value has been found, but $D \gtrsim 45$ Mpc, allowing the point-like approximation ($\theta_{\mathrm{HCN}} = 0$) with an error $\lesssim 5\%$ if the dense molecular gas in this galaxy is not unusually widely distributed. For galaxies marked with a dash (-), no value has been found, and $D < 45$ Mpc.}
 	 \tablefoottext{f}{From \citet{ned}.}
 	 \tablefoottext{g}{\object{IC 694} and \object{NGC 3690} are together also known as the merger \object{Arp 299}.}
 	 \tablefoottext{h}{This is the FIR luminosity of \object{IC 694} and \object{NGC 3690} together.}
 	 \tablefoottext{i}{From \citet{sanders03}.}
 	 \tablefoottext{j}{Short for \object{IRAS 17208-0014}.}
 	 \tablefoottext{k}{For the HCN~1\nobrkhyph0 data the following coordinates have been used: R.A. 12 26 54.63, Decl. --00 52 39.6 (J2000.0).}
 	 }    
\end{table*}

\section{Results}
\label{sec:res}

\begin{figure*}[hb]
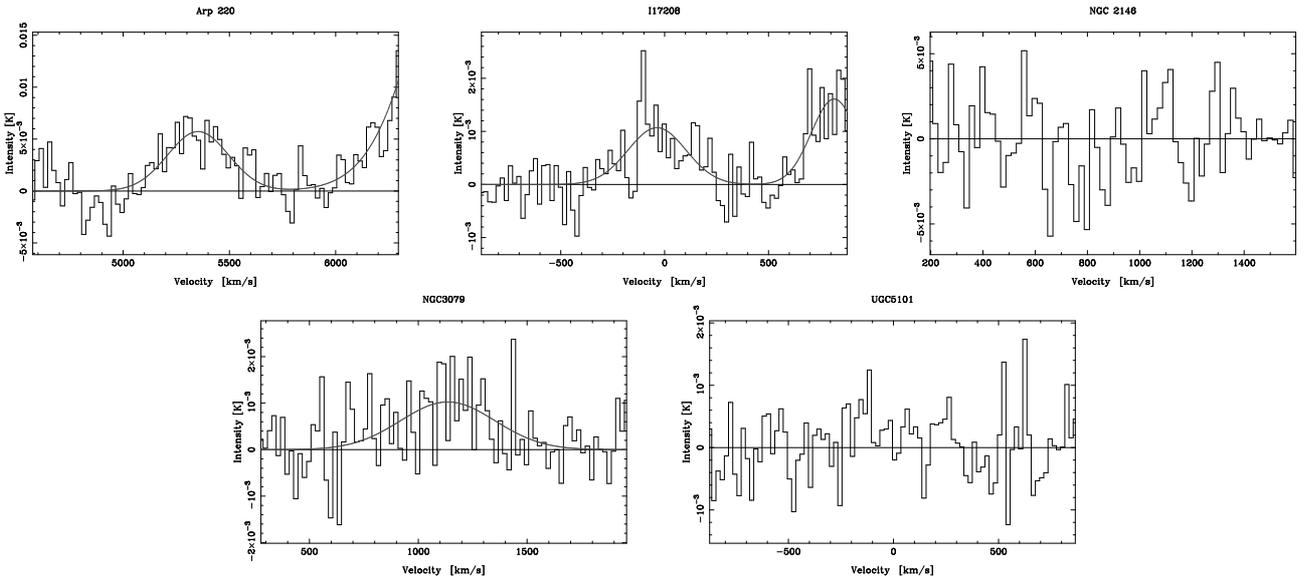

   \centering
   $\begin{array}{c@{\hspace{0.7cm}}c@{\hspace{0.7cm}}c}
	\includegraphics[width=5.2cm]{15565fg6a.eps}  &  

	\includegraphics[width=5.2cm]{15565fg6b.eps}  &
	\includegraphics[width=5.2cm]{15565fg6c.eps} \\
	&
	\hspace{-5.8cm}
	\includegraphics[width=5.2cm]{15565fg6d.eps}  &  
	\hspace{-5.8cm}
	\includegraphics[width=5.2cm]{15565fg6e.eps} \\
	\end{array}$
   \caption{HC$_3$N~10\nobrkhyph9 spectra for \object{Arp 220}, \object{IRAS 17208-0014} (detections), \object{NGC 2146} (non-detection), \object{NGC 3079} (detection), and \object{UGC 5101} (non-detection). For \object{Arp 220} and \object{IRAS 17208-0014}, part of the HNC~1\nobrkhyph0 line is also visible. Some of the reported HC$_3$N~10\nobrkhyph9-detections are found in the HNC~1\nobrkhyph0 spectra, see Figure~\ref{fig:hnc1-0}.}
              \label{fig:hc3n10-9}%
    \end{figure*}

\begin{figure*}
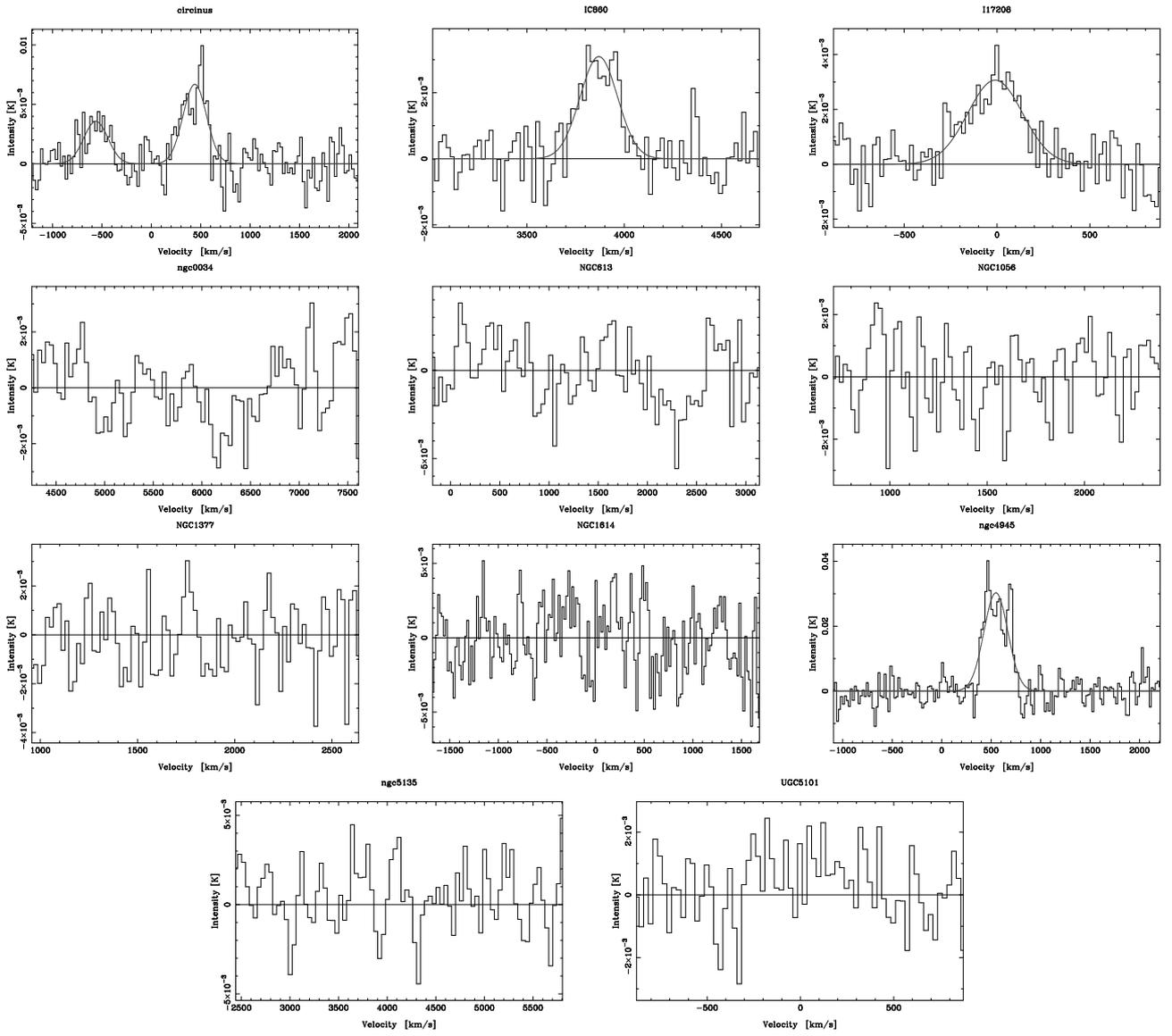

   \centering
   $\begin{array}{c@{\hspace{0.7cm}}c@{\hspace{0.7cm}}c}
	\includegraphics[width=5.2cm]{15565fg12a.eps}  &  
	\includegraphics[width=5.2cm]{15565fg12b.eps} &
	\includegraphics[width=5.2cm]{15565fg12c.eps} \\
	\includegraphics[width=5.2cm]{15565fg12d.eps} &
	\includegraphics[width=5.2cm]{15565fg12e.eps} & 
	\includegraphics[width=5.2cm]{15565fg12f.eps} \\
	\includegraphics[width=5.2cm]{15565fg12g.eps}  &
	\includegraphics[width=5.2cm]{15565fg12h.eps} & 
	\includegraphics[width=5.2cm]{15565fg12i.eps} \\
	&
	\hspace{-5.8cm}
	\includegraphics[width=5.2cm]{15565fg12j.eps} &
	\hspace{-5.8cm}
	\includegraphics[width=5.2cm]{15565fg12k.eps} \\
	\end{array}$
	\caption{HNC~1\nobrkhyph0 spectra for \object{Circinus}, \object{IC 860}, \object{IRAS 17208-0014} (detections), \object{NGC 34}, \object{NGC 613}, \object{NGC 1056}, \object{NGC 1377}, \object{NGC 1614} (non-detections), \object{NGC 4945} (detection), \object{NGC 5135}, and \object{UGC 5101} (non-detections). For \object{Circinus}, the HC$_3$N~10\nobrkhyph9 line is also visible. Bandwidths for the spectra of \object{NGC 34}, \object{NGC 613}, \object{NGC 1614}, and \object{NGC 5135} are broad enough to include non-detections of the HC$_3$N~10\nobrkhyph9 line, thus giving upper limits for these lines.}
              \label{fig:hnc1-0}%
    \end{figure*}

All new HC$_3$N, HCN, and HNC spectra reported in this work are displayed in Figures~\ref{fig:hc3n10-9}\nobrkhyph\ref{fig:hnc1-0}. The observed values of the spectral line intensities can be found in Tables~\ref{tab:resultshc3n}\nobrkhyph\ref{tab:resultshnc}. Data from the literature are also included in these tables.

Some observations of HC$_3$N~10\nobrkhyph9 and HNC~1\nobrkhyph0 performed with SEST include both these lines in the same spectrum, due to the large bandwidth. The spectra are in these cases labelled according to the central peak. The frequency difference between the two peaks is 315~MHz, as can be seen in Table~\ref{tab:obsparam}. This gives a velocity difference of approximately 1000~km/s.

\subsection{New detections}

This is not only the first text to put together a survey of all extragalactic HC$_3$N emission line data, but it also reports the first HC$_3$N detections in six galaxies: \object{Circinus}, \object{IC 860}, \object{IRAS 17208-0014}, \object{Maffei 2}, \object{NGC 1068}, and \object{NGC 3079}. The number of extragalactic sources where HC$_3$N has been detected is thus almost doubled. Three of the HNC detections are also made in sources without earlier HNC detections: \object{Circinus}, \object{IC 860}, and \object{IRAS 17208-0014}. Finally, the first detection of HCN in \object{NGC 613} is also reported.

\begin{table*}
\centering
      \caption[]{Data from HC$_3$N observations.}
         \label{tab:resultshc3n}
         \begin{tabular}{l l c c c l c c l l}
 
            \noalign{\smallskip}
			\hline
            \hline
            \noalign{\smallskip}
            Galaxy & Line & $I$(HC$_3$N)\tablefootmark{a} & $S_{\nu} \Delta v$ & $\Delta v$ & Telescope & $\eta_{\mathrm{mb}}$ & $\theta_{\mathrm{mb}}$ & $T$ scale\tablefootmark{a} & References \\
            & & [K km s$^{-1}$] & [Jy km s$^{-1}$] & [km s$^{-1}$] & & & [$''$] & & \\
            \noalign{\smallskip}
            \hline
            \noalign{\smallskip}
			\object{Arp 220} & 10-9 & $2.02 \pm 0.15$ & 45 $\pm$ 3 & 340 & OSO 20 m & 0.59 & 44 & $T_{\mathrm{A}}^*$ & (1) \\
			\object{Arp 220} & 10-9\tablefootmark{b} & $0.4 \pm 0.15$ & 11 $\pm$ 4 & 350 & SEST 15 m & 0.75 & 55 & $T_{\mathrm{A}}^*$ & (2) \\
			\object{Arp 220} & 12-11 & $0.96 \pm 0.1$ & 23 $\pm$ 2 & 170 & OSO 20 m & 0.52 & 36 & $T_{\mathrm{A}}^*$ & (1) \\
			\object{Circinus} & 10-9\tablefootmark{b} & $1.01 \pm 0.1$ & 27.6 $\pm$ 2.7 & 290 & SEST 15 m & 0.75 & 55 & $T_{\mathrm{A}}^*$ & (1) \\
			\object{IC 342} & 10-9 & $2.6 \pm 0.7$ & 14 $\pm$ 4 & 52 & IRAM 30 m & 0.8 & 25 & $T_{\mathrm{R}}^*$ & (3) \\
			\object{IC 694} & 12-11 & $<0.30$ & $<$7.4 & ... & OSO 20 m & 0.52 & 36 & $T_{\mathrm{A}}^*$ & (1) \\
			\object{IC 860} & 28-27 & $0.54 \pm 0.07$ & 3.9 $\pm$ 0.5 & 175 & IRAM 30 m & 0.59 & 9 & $T_{\mathrm{A}}^*$ & (1) \\
			I17208 & 10-9 & $0.33 \pm 0.03$ & 2.2 $\pm$ 0.2 & 330 & IRAM 30 m & 0.80 & 28 & $T_{\mathrm{A}}^*$ & (1) \\
			\object{M82} & 12-11 & $5.6 \pm 0.6$ & 30 $\pm$ 3 & 155 & IRAM 30 m & 0.80 & 25 & $T_{\mathrm{R}}^*$ & (3) \\
			\object{Maffei 2} & 12-11 & $1.42 \pm 0.05$ & 10.9 $\pm$ 0.4 & 200 & IRAM 30 m & 0.73 & 24 & $T_{\mathrm{A}}^*$ & (1) \\
			\object{NGC 34} & 10-9\tablefootmark{b} & $<0.45$ & $<$12 & ... & SEST 15 m & 0.75 & 55 & $T_{\mathrm{A}}^*$ & (1) \\
			\object{NGC 253} & 9-8 & $5.8 \pm 0.6$ & 27 $\pm$ 3 & 63\tablefootmark{c} & IRAM 30 m & ... & 29 & $T_{\mathrm{mb}}$ & (4) \\
			\object{NGC 253} & 10-9 & $5.8 \pm 0.6$ & 27 $\pm$ 3 & 63\tablefootmark{c} & IRAM 30 m & ... & 26 & $T_{\mathrm{mb}}$ & (4) \\
			\object{NGC 253} & 12-11 & $4.4 \pm 0.7$ & 19 $\pm$ 3 & 63\tablefootmark{c} & IRAM 30 m & ... & 21 & $T_{\mathrm{mb}}$ & (4) \\
			\object{NGC 253} & 15-14 & $4.4 \pm 0.3$ & 24 $\pm$ 2 & 77, 85 & IRAM 30 m & ... & 19 & $T_{\mathrm{mb}}$ & (5) \\
			\object{NGC 253} & 15-14 & $3.6 \pm 0.6$ & 16 $\pm$ 3 & 63\tablefootmark{c} & IRAM 30 m & ... & 17 & $T_{\mathrm{mb}}$ & (4) \\
			\object{NGC 253} & 16-15 & $3.8$ & 19 & 77\tablefootmark{c}, 85\tablefootmark{c} & IRAM 30 m & ... & 17 & $T_{\mathrm{mb}}$ & (5) \\
			\object{NGC 253} & 17-16 & $3.0 \pm 0.2$ & 15 $\pm$ 1 & 72\tablefootmark{c} & IRAM 30 m & ... & 16 & $T_{\mathrm{mb}}$ & (5) \\
			\object{NGC 253} & 17-16 & $3.4 \pm 0.4$ & 15 $\pm$ 2 & 63\tablefootmark{c} & IRAM 30 m & ... & 15 & $T_{\mathrm{mb}}$ & (4) \\
			\object{NGC 253} & 18-17 & $2.2 \pm 0.5$ & 11 $\pm$ 2 & 73 & IRAM 30 m & ... & 15 & $T_{\mathrm{mb}}$ & (5) \\
			\object{NGC 253} & 19-18 & $4.6 \pm 0.6$ & 22 $\pm$ 3 & 74 & IRAM 30 m & ... & 14 & $T_{\mathrm{mb}}$ & (5) \\
			\object{NGC 253} & 26-25 & $3.2 \pm 0.7$ & 21 $\pm$ 5 & 63\tablefootmark{c} & IRAM 30 m & ... & 12 & $T_{\mathrm{mb}}$ & (4) \\
			\object{NGC 613} & 10-9\tablefootmark{b} & $<0.26$ & $<$7.0 & ... & SEST 15 m & 0.75 & 55 & $T_{\mathrm{A}}^*$ & (1) \\
			\object{NGC 1056} & 16-15 & $<0.17$ & $<$1.3 & ... & IRAM 30 m & 0.67 & 17 & $T_{\mathrm{A}}^*$ & (1) \\
			\object{NGC 1068} & 10-9\tablefootmark{b} & $0.39 \pm 0.05$ & 11 $\pm$ 1 & 100 & SEST 15 m & 0.75 & 55 & $T_{\mathrm{A}}^*$ & (1)\tablefootmark{d} \\
			\object{NGC 1365} & 10-9\tablefootmark{b} & $<0.61$ & $<$417 & ... & SEST 15 m & 0.75 & 55 & $T_{\mathrm{A}}^*$ & (1)\tablefootmark{d} \\
			\object{NGC 1377} & 16-15 & $<0.28$ & $<$2.1 & ... & IRAM 30 m & 0.67 & 17 & $T_{\mathrm{A}}^*$ & (1) \\
			\object{NGC 1377} & 25-24 & $<0.26$ & $<$1.9 & ... & IRAM 30 m & 0.63 & 10.5 & $T_{\mathrm{A}}^*$ & (1) \\
			\object{NGC 1614} & 10-9\tablefootmark{b} & $<0.39$ & $<$11 & ... & SEST 15 m & 0.75 & 55 & $T_{\mathrm{A}}^*$ & (1) \\
			\object{NGC 1808} & 10-9\tablefootmark{b} & $0.2 \pm 0.1$ & 6 $\pm$ 3 & 250 & SEST 15 m & 0.75 & 57 & $T_{\mathrm{A}}^*$ & (2) \\
			\object{NGC 2146} & 10-9 & $<0.38$ & $<$7.7 & ... & OSO 20 m & 0.59 & 42 & $T_{\mathrm{A}}^*$ & (1) \\
			\object{NGC 2146} & 12-11 & $<0.38$ & $<$9.2 & ... & OSO 20 m & 0.52 & 36 & $T_{\mathrm{A}}^*$ & (1) \\
			\object{NGC 2623} & 12-11 & $<0.50$ & $<$12 & ... & OSO 20 m & 0.52 & 36 & $T_{\mathrm{A}}^*$ & (1) \\
			\object{NGC 3079} & 10-9 & $0.60 \pm$ 0.05 & 4.0 $\pm$ 0.3 & 500 & IRAM 30 m & 0.80 & 28 & $T_{\mathrm{A}}^*$ & (1) \\
			\object{NGC 3079} & 16-15 & $<0.54$ & $<$4 & ... & IRAM 30 m & 0.80 & 17 & $T_{\mathrm{A}}^*$ & (1) \\
			\object{NGC 3079} & 25-24 & $<0.40$ & $<$3 & ... & IRAM 30 m & 0.80 & 10.5 & $T_{\mathrm{A}}^*$ & (1) \\
			\object{NGC 3256} & 10-9\tablefootmark{b} & $<0.12$ & $<$3.3 & ... & SEST 15 m & 0.75 & 55 & $T_{\mathrm{A}}^*$ & (2) \\
			\object{NGC 3690} & 12-11 & $<0.31$ & $<$7.5 & ... & OSO 20 m & 0.52 & 36 & $T_{\mathrm{A}}^*$ & (1) \\
			\object{NGC 4418}\tablefootmark{e} & 10-9 & $0.8 \pm 0.08$ & 5 $\pm$ 0.5 & 122 & IRAM 30 m & 0.77 & 27 & $T_{\mathrm{A}}^*$ & (6) \\
			\object{NGC 4418}\tablefootmark{e} & 16-15\tablefootmark{f} & $1.7 \pm 0.08$ & 12 $\pm$ 0.6 & 130 & IRAM 30 m & 0.70 & 17 & $T_{\mathrm{A}}^*$ & (6) \\
			\object{NGC 4418}\tablefootmark{e} & 25-24 & $1.6 \pm 0.2$ & 15 $\pm$ 2 & 140 & IRAM 30 m & 0.53 & 11 & $T_{\mathrm{A}}^*$ & (6) \\
			\object{NGC 4945} & 9-8 & $2.16 \pm 0.50$ & 47 $\pm$ 11 & 230 & SEST 15 m & 0.78 & 63 & $T_{\mathrm{mb}}$ & (7) \\
			\object{NGC 4945} & 10-9 & $1.99 \pm 0.21$ & 41 $\pm$ 4 & 290 & SEST 15 m & 0.75 & 55 & $T_{\mathrm{mb}}$ & (7) \\
			\object{NGC 4945} & 11-10 & $2.92 \pm 0.35$ & 65 $\pm$ 8 & 340 & SEST 15 m & 0.73 & 52 & $T_{\mathrm{mb}}$ & (7) \\
			\object{NGC 4945} & 12-11 & $4.18 \pm 0.38$ & 98 $\pm$ 9 & 340 & SEST 15 m & 0.71 & 49 & $T_{\mathrm{mb}}$ & (7) \\
			\object{NGC 4945} & 15-14 & $2.13 \pm 0.29$ & 52 $\pm$ 7 & 250 & SEST 15 m & 0.65 & 40 & $T_{\mathrm{mb}}$ & (7) \\
			\object{NGC 4945} & 16-15\tablefootmark{g} & $5.02 \pm 0.19$ & 120 $\pm$ 5 & 330 & SEST 15 m & 0.63 & 37 & $T_{\mathrm{mb}}$ & (7) \\
			\object{NGC 4945} & 17-16 & $2.26 \pm 0.55$ & 48 $\pm$ 12 & 280 & SEST 15 m & 0.61 & 33 & $T_{\mathrm{mb}}$ & (7) \\
			\object{NGC 4945} & 24-23 & $<0.60$ & $<$12 & ... & SEST 15 m & 0.48 & 23 & $T_{\mathrm{mb}}$ & (7) \\
			\object{NGC 4945} & 25-24 & $<0.60$ & $<$12 & ... & SEST 15 m & 0.46 & 22 & $T_{\mathrm{mb}}$ & (7) \\
			\object{NGC 5135} & 10-9\tablefootmark{b} & $<0.13$ & $<$3.7 & ... & SEST 15 m & 0.75 & 55 & $T_{\mathrm{A}}^*$ & (1) \\
			\object{NGC 6946} & 12-11 & $<0.32$ & $<$2.5 & ... & IRAM 30 m & 0.73 & 24 & $T_{\mathrm{A}}^*$ & (1) \\
			\object{NGC 7130} & 10-9\tablefootmark{b} & $<0.10$ & $<$2.7 & ... & SEST 15 m & 0.75 & 55 & $T_{\mathrm{A}}^*$ & (2) \\
			\object{UGC 5101} & 10-9 & $<0.12$ & $<$0.76 & ... & IRAM 30 m & 0.80 & 28 & $T_{\mathrm{A}}^*$ & (1) \\
            \noalign{\smallskip}
            \hline

     \end{tabular}

\tablebib{(1) This work; (2) \citet{aalto02}; (3) \citet{henkel88}; (4) \citet{mauersberger90}; (5) \citet{martin06}; (6) \citet{aalto07a}; (7) \citet{wang04}.}
\tablefoot{
\tablefoottext{a}{The temperature scale of the integrated intensities are given in the $T$ scale column. Upper limits are $2\sigma$ calculated from the rms of the noise surrounding the line for our data. For our data, errors are given in 1$\sigma$ and calculated from the rms. Since many articles lack information about sizes of errors and methods used when calculating the errors, most errors are given as printed in the respective article. However, if the size of the error is clearly written, it has been recalculated to 1$\sigma$.}
\tablefoottext{b}{Measured in HNC~1\nobrkhyph0 spectrum.}
\tablefoottext{c}{Fixed when fitting Gaussian.}
\tablefoottext{d}{These values are calculated from HNC~1\nobrkhyph0 spectra already published in \citet{perez07}.}
\tablefoottext{e}{See \citet{costagliola10a} for an extensive survey of HC$_3$N in this galaxy, with data not included in this table.}
\tablefoottext{f}{Contaminated by para-H$_2$CO, estimated to $20\%$. The given value is only for the HC$_3$N component.}
\tablefoottext{g}{Contaminated by para-H$_2$CO.}
}
\end{table*}

\begin{table*}
\centering
      \caption[]{Data from HCN~1\nobrkhyph0 observations}
         \label{tab:resultshcn}
         \begin{tabular}{l c c c l c c l l}
 
            \noalign{\smallskip}
			\hline
            \hline
            \noalign{\smallskip}
            Galaxy & $I$(HCN) 1-0\tablefootmark{a} & $S_{\nu} \Delta v$ & $\Delta v$ & Telescope & $\eta_{\mathrm{mb}}$ & $\theta_{\mathrm{mb}}$ & $T$ scale\tablefootmark{a} & References \\
            & [K km s$^{-1}$] & [Jy km s$^{-1}$] & [km s$^{-1}$] & & & [$''$] & & \\
            \noalign{\smallskip}
            \hline
            \noalign{\smallskip}
			\object{Arp 220} & $9.7 \pm 0.4$ & 57 $\pm$ 2 & 530 & IRAM 30 m & 0.82 & 29.5 & $T_{\mathrm{mb}}$ & (1) \\
			\object{Circinus} & $5.2 \pm 0.8$ & 110 $\pm$ 20 & 300 & SEST 15 m & 0.75 & 57 & $T_{\mathrm{mb}}$ & (2) \\
			\object{IC 342} & $15.5 \pm 0.9$ & 36 $\pm$ 2 & ... & NRO 45 m & 0.54 & 19 & $T_{\mathrm{mb}}$ & (3) \\
			\object{IC 694} & $1.29 \pm 0.09$ & 8.4 $\pm$ 0.6 & ... & IRAM 30 m & 0.82 & 28 & $T_{\mathrm{A}}^*$ & (4) \\
			\object{IC 860} & ... & ... & ... & & ... & ... & & N/A\tablefootmark{b} \\
			I17208 & $2.19 \pm 0.16$ & 14 $\pm$ 1 & ... & IRAM 30 m & 0.82 & 28 & $T_{\mathrm{A}}^*$ & (4) \\
			I17208 & $0.91 \pm 0.19$ & 34 $\pm$ 7 & ... & NRAO 12 m & 0.89 & 72 & $T_{\mathrm{R}}^*$ & (5) \\
			\object{M82} & $29 \pm 0.2$ & 170 $\pm$ 1 & 130 & IRAM 30 m & 0.82 & 29.5 & $T_{\mathrm{mb}}$ & (1) \\
			\object{Maffei 2} & $13.8 \pm 0.9$ & 32 $\pm$ 2 & ... & NRO 45 m & 0.54 & 19 & $T_{\mathrm{mb}}$ & (3) \\
			\object{NGC 34} & $1.6 \pm 0.2$ & 33 $\pm$ 4 & 600-700 & SEST 15 m & 0.75 & 57 & $T_{\mathrm{mb}}$ & (6) \\
			\object{NGC 253} & $40.8$ & 175 & 150 & NRO 45 m & 0.45 & 23 & $T_{\mathrm{A}}^*$ & (7) \\
			\object{NGC 613} & $0.53 \pm 0.08$ & 15 $\pm$ 2 & 130 & SEST 15 m & 0.75 & 57 & $T_{\mathrm{A}}^*$ & (8) \\
			\object{NGC 1056} & ... & ... & ... & & ... & ... & & N/A\tablefootmark{b} \\
			\object{NGC 1068} & $24.5 \pm 0.9$ & 145 $\pm$ 5 & 220 & IRAM 30 m & 0.82 & 29.5 & $T_{\mathrm{mb}}$ & (1) \\
			\object{NGC 1365} & $6.0 \pm 0.1$ & 125 $\pm$ 2 & 300-400 & SEST 15 m & 0.75 & 57 & $T_{\mathrm{mb}}$ & (6) \\
			\object{NGC 1377} & $0.47 \pm 0.1$ & 3.0 $\pm$ 0.6 & 140 & IRAM 30 m & 0.80 & 28 & $T_{\mathrm{A}}^*$ & (8) \\
			\object{NGC 1614} & $1.5 \pm 0.22$ & 40 $\pm$ 6 & 300 & FCRAO 14 m & 0.60 & 50 & $T_{\mathrm{A}}^*$ & (5) \\
			\object{NGC 1808} & $4$ & 110 & & SEST 15 m & 0.74 & 56 & $T_{\mathrm{A}}^*$ & (9) \\
			\object{NGC 2146} & $5 \pm 0.1$ & 30 $\pm$ 1 & 290 & IRAM 30 m & 0.82 & 29.5 & $T_{\mathrm{mb}}$ & (1) \\
			\object{NGC 2623} & ... & ... & ... &  & ... & ... & & N/A\tablefootmark{b} \\
			\object{NGC 3079} & $5.7\pm 0.8$ & 29 $\pm$ 4 & 420 & IRAM 30 m & 0.80 & 28 & $T_{\mathrm{mb}}$& (10) \\
			\object{NGC 3079} & $2.6 \pm 0.42$ & 97 $\pm$ 16 & 365 & NRAO 12 m & 0.89 & 72 & $T_{\mathrm{R}}^*$ & (5) \\
			\object{NGC 3256} & $2.3 \pm 0.2$ & 48 $\pm$ 4 & 165 & SEST 15 m & 0.77 & 57 & $T_{\mathrm{mb}}$ & (11) \\
			\object{NGC 3690} & $2.04 \pm 0.11$ & 13.2 $\pm$ 0.7 & 300 & IRAM 30 m & 0.82 & 28 & $T_{\mathrm{A}}^*$ & (4) \\
			\object{NGC 4418} & $1.96 \pm 0.04$ & 12.4 $\pm$ 0.3 & 170 & IRAM 30 m & 0.80 & 28 & $T_{\mathrm{A}}^*$ & (8) \\
			\object{NGC 4945} & $22.4 \pm 0.4$ & 436 $\pm$ 8 & 305 & SEST 15 m & 0.75 & 55 & $T_{\mathrm{mb}}$ & (12) \\
			\object{NGC 5135} & $0.65 \pm 0.07$ & 14 $\pm$ 1.5 & 50-60 & SEST 15 m & 0.75 & 57 & $T_{\mathrm{mb}}$ & (6) \\
			\object{NGC 6946} & $8.7 \pm 0.9$ & 20 $\pm$ 2 & ... & NRO 45 m & 0.54 & 19 & $T_{\mathrm{mb}}$ & (3) \\
			\object{NGC 7130} & $0.7 \pm 0.1$ & 15 $\pm$ 2 & 100 & SEST 15 m & 0.75 & 57 & $T_{\mathrm{mb}}$ & (6) \\
			\object{UGC 5101} & $1.40 \pm 0.14$ & 9.1 $\pm$ 0.9 & 500 & IRAM 30 m & 0.82 & 28 & $T_{\mathrm{A}}^*$ & (4) \\
            \noalign{\smallskip}
            \hline

     \end{tabular}
     
     \tablebib{(1) \citet{krips08}; (2) \citet{curran01a}; (3) \citet{sorai02}; (4) \citet{gracia08}; (5) \citet{gao04}; (6) \citet{curran00}; (7) \citet{nguyen89}; (8) This work; (9) \citet{aalto94}; (10) \citet{perez07}; (11) \citet{casoli92}; (12) \citet{wang04}.}
     \tablefoot{
     \tablefoottext{a}{The temperature scale of the integrated intensities are given in the $T$ scale column. Upper limits are $2\sigma$ calculated from the rms of the noise surrounding the line for our data. For our data, errors are given in 1$\sigma$ and calculated from the rms. Since many articles lack information about sizes of errors and methods used when calculating the errors, most errors are given as printed in the respective article. However, if the size of the error is clearly written, it has been recalculated to 1$\sigma$.}
     \tablefoottext{b}{No single-dish HCN data were found in the literature for these objects.}
	}
\end{table*}

\begin{table*}
\centering
      \caption[]{Data from HNC~1\nobrkhyph0 observations}
         \label{tab:resultshnc}
         \begin{tabular}{l c c c l c c l l}
 
            \noalign{\smallskip}
			\hline
            \hline
            \noalign{\smallskip}
            Galaxy & $I$(HNC) 1-0\tablefootmark{a} & $S_{\nu} \Delta v$ & $\Delta v$ & Telescope & $\eta_{\mathrm{mb}}$ & $\theta_{\mathrm{mb}}$ & $T$ scale\tablefootmark{a} & References \\
            & [K km s$^{-1}$] & [Jy km s$^{-1}$] & [km s$^{-1}$] & & & [$''$] & & \\
            \noalign{\smallskip}
            \hline
            \noalign{\smallskip}
			\object{Arp 220} & $0.95 \pm 0.2$ & 26 $\pm$ 6 & ... & SEST 15 m & 0.75 & 55 & $T_{\mathrm{A}}^*$ & (1) \\
			\object{Circinus} & $1.99 \pm 0.1$ & 55 $\pm$ 3 & 280 & SEST 15 m & 0.75 & 55 & $T_{\mathrm{A}}^*$ & (2) \\
			\object{IC 342} & $9.2 \pm 0.7$ & 48 $\pm$ 4 & 47 & IRAM 30 m & 0.8 & 25 & $T_{\mathrm{R}}^*$ & (3) \\
			\object{IC 694} & $0.75 \pm 0.2$ & 15 $\pm$ 4 & 300-400 & OSO 20 m & 0.59 & 42 & $T_{\mathrm{A}}^*$ & (1) \\
			\object{IC 860} & $0.70 \pm 0.04$ & 4.6 $\pm$ 0.3 & 230 & IRAM 30 m & 0.80 & 28 & $T_{\mathrm{A}}^*$ & (2) \\
			I17208 & $1.12 \pm 0.06$ & 7.4 $\pm$ 0.4 & 350 & IRAM 30 m & 0.80 & 28 & $T_{\mathrm{A}}^*$ & (2) \\
			\object{M82} & $7.3 \pm 0.6$ & 31 $\pm$ 3 & 129 & IRAM 30 m & 0.64 & 25 & $T_{\mathrm{mb}}$ & (4) \\
			\object{Maffei 2} & $7.3 \pm 0.8$ & 31 $\pm$ 3 & 91, 50 & IRAM 30 m & 0.64 & 25 & $T_{\mathrm{mb}}$ & (4) \\
			\object{NGC 34} & $<0.44$ & $<$12 & ... & SEST 15 m & 0.75 & 55 & $T_{\mathrm{A}}^*$ & (2) \\
			\object{NGC 253} & $50.0 \pm 2.8$ & 210 $\pm$ 12 & 72, 136 & IRAM 30 m & 0.64 & 25 & $T_{\mathrm{mb}}$ & (4) \\
			\object{NGC 613} & $<0.24$ & $<$6.5 & ... & SEST 15 m & 0.75 & 55 & $T_{\mathrm{A}}^*$ & (2) \\
			\object{NGC 1056} & $<0.17$ & $<$1.1 & ... & IRAM 30 m & 0.80 & 28 & $T_{\mathrm{A}}^*$ & (2) \\
			\object{NGC 1068} & $3.2 \pm 0.5$ & 65 $\pm$ 10 & 260 & SEST 15 m & 0.75 & 55 & $T_{\mathrm{mb}}$ & (5) \\
			\object{NGC 1068} & $11.4 \pm 0.7$ & 48 $\pm$ 3 & 232 & IRAM 30 m & 0.64 & 25 & $T_{\mathrm{mb}}$ & (4) \\
			\object{NGC 1365} & $4.7 \pm 0.6$ & 96 $\pm$ 12 & 150 & SEST 15 m & 0.75 & 55 & $T_{\mathrm{mb}}$ & (5) \\
			\object{NGC 1377} & $<0.15$ & $<$0.99 & ... & IRAM 30 m & 0.80 & 28 & $T_{\mathrm{A}}^*$ & (2) \\
			\object{NGC 1614} & $<0.38$ & $<$10 & ... & SEST 15 m & 0.75 & 55 & $T_{\mathrm{A}}^*$ & (2) \\
			\object{NGC 1808} & $1.2 \pm 0.1$ & 33 $\pm$ 3 & 300 & SEST 15 m & 0.75 & 55 & $T_{\mathrm{A}}^*$ & (1) \\
			\object{NGC 2146} & $1.6 \pm 0.3$ & 6.7 $\pm$ 1.3 & 237 & IRAM 30 m & 0.64 & 25 & $T_{\mathrm{mb}}$ & (4) \\
			\object{NGC 2623} & $0.6 \pm 0.15$ & 12 $\pm$ 3 & 500-600 & OSO 20 m & 0.59 & 42 & $T_{\mathrm{A}}^*$ & (1) \\
			\object{NGC 3079} & $2.9 \pm 0.5$ & 15 $\pm$ 3 & 380 & IRAM 30 m & 0.80 & 28 & $T_{\mathrm{mb}}$ & (5) \\
			\object{NGC 3079} & $6.9 \pm 1.0$ & 29 $\pm$ 4 & 545 & IRAM 30 m & 0.64 & 25 & $T_{\mathrm{mb}}$ & (4) \\
			\object{NGC 3256} & $0.6 \pm 0.05$ & 16 $\pm$ 1 & 250 & SEST 15 m & 0.75 & 55 & $T_{\mathrm{A}}^*$ & (1) \\
			\object{NGC 3690} & ... & ... & ... & & ... & ... & & N/A\tablefootmark{b} \\
			\object{NGC 4418} & $1.24 \pm 0.12$ & 7.9 $\pm$ 0.8 & 156 & IRAM 30 m & 0.77 & 27 & $T_{\mathrm{A}}^*$ & (6) \\
			\object{NGC 4945} & $8.6 \pm 0.2$ & 230 $\pm$ 5 & 290 & SEST 15 m & 0.75 & 55 & $T_{\mathrm{A}}^*$ & (2) \\
			\object{NGC 5135} & $<0.13$ & $<$3.5 & ... & SEST 15 m & 0.75 & 55 & $T_{\mathrm{A}}^*$ & (2) \\
			\object{NGC 6946} & $4.0 \pm 0.3$ & 17 $\pm$ 1 & 138 & IRAM 30 m & 0.64 & 25 & $T_{\mathrm{mb}}$ & (4) \\
			\object{NGC 7130} & $0.4 \pm 0.05$ & 11 $\pm$ 1 & ... & SEST 15 m & 0.75 & 55 & $T_{\mathrm{A}}^*$ & (1) \\
			\object{UGC 5101} & $1.24 \pm 0.1$ & 8.2 $\pm$ 0.7 & 500 & IRAM 30 m & 0.80 & 28 & $T_{\mathrm{A}}^*$ & (2) \\
            \noalign{\smallskip}
            \hline

     \end{tabular}
	 \tablebib{(1) \citet{aalto02}; (2) This work; (3) \citet{henkel88}; (4) \citet{huttemeister95}; (5) \citet{perez07}; (6) \citet{aalto07a}.}
     \tablefoot{
     \tablefoottext{a}{The temperature scale of the integrated intensities are given in the $T$ scale column. Upper limits are $2\sigma$ calculated from the rms of the noise surrounding the line for our data. For our data, errors are given in 1$\sigma$ and calculated from the rms. Since many articles lack information about sizes of errors and methods used when calculating the errors, most errors are given as printed in the respective article. However, if the size of the error is clearly written, it has been recalculated to 1$\sigma$.}
     \tablefoottext{b}{No HNC data were found in the literature for this object.}
	} 
\end{table*}

\subsection{Line ratios}

\begin{table}[H!]
\centering
      \caption[]{Calculated line ratios.}
         \label{tab:ratios}
         \begin{tabular}{l c c c}
 
            \noalign{\smallskip}
			\hline
            \hline
            \noalign{\smallskip}
            Galaxy & $\frac{I(\mathrm{HC}_3\mathrm{N})}{I(\mathrm{HCN}\:1-0)}$ & $\frac{I(\mathrm{HC}_3\mathrm{N})}{I(\mathrm{HNC}\:1-0)}$ & $\frac{I(\mathrm{HNC}\:1-0)}{I(\mathrm{HCN}\:1-0)}$ \\
            \noalign{\smallskip}
            \hline
            \noalign{\smallskip}
			\object{Arp 220} & $0.19 \pm 0.07$\tablefootmark{a} & $0.42 \pm 0.2$\tablefootmark{a} & $0.45 \pm 0.1$ \\
			\object{Arp 220} & $0.78 \pm 0.06$\tablefootmark{b} & $1.73 \pm 0.4$\tablefootmark{b} & $0.45 \pm 0.1$ \\
			\object{Circinus} & $0.24 \pm 0.04$ & $0.51 \pm 0.06$ & $0.48 \pm 0.08$ \\
			\object{IC 342} & $0.28 \pm 0.08$ & $0.28 \pm 0.08$ & $1.00 \pm 0.1$ \\
			\object{IC 694} & $<0.61$\tablefootmark{c} & $<0.34$\tablefootmark{c} & $1.79 \pm 0.5$ \\
			\object{IC 860} & ... & $0.11 \pm 0.02$\tablefootmark{d} & ... \\
			I17208 & $0.16 \pm 0.02$\tablefootmark{e} & $0.30 \pm 0.03$ & $0.53 \pm 0.05$\tablefootmark{e} \\
			I17208 & $0.062 \pm 0.01$\tablefootmark{f} & $0.30 \pm 0.03$ & $0.21 \pm 0.05$\tablefootmark{f} \\
			\object{M82} & $0.21 \pm 0.02$\tablefootmark{c} & $0.96 \pm 0.1$\tablefootmark{c} & $0.22 \pm 0.02$ \\
			\object{Maffei 2} & $0.20 \pm 0.02$\tablefootmark{c} & $0.25 \pm 0.03$\tablefootmark{c} & $0.81 \pm 0.1$ \\
			\object{NGC 34} & $<0.35$ & ... & $<0.34$ \\
			\object{NGC 253} & $0.094 \pm 0.01$ & $0.15 \pm 0.02$ & $0.63 \pm 0.04$ \\
			\object{NGC 613} & $<0.46$ & ... & $<0.44$ \\
			\object{NGC 1056} & ... & ... & ... \\
			\object{NGC 1068} & $0.072 \pm 0.06$ & $0.15 \pm 0.15$\tablefootmark{g} & $0.45 \pm 0.19$\tablefootmark{g} \\
			\object{NGC 1365} & $<0.13$ & $<0.17$ & $0.74 \pm 0.20$ \\
			\object{NGC 1377} & $<0.26$\tablefootmark{h} & ... & $<0.32$ \\			
			\object{NGC 1614} & $<0.25$ & ... & $<0.24$ \\
			\object{NGC 1808} & $0.05 \pm 0.025$ & $0.18 \pm 0.09$ & $ 0.29 \pm 0.02$ \\
			\object{NGC 2146} & $<0.20$ & $<0.85$ & $0.26 \pm 0.06$ \\
			\object{NGC 2623} & ... & $<0.70$\tablefootmark{c} & ... \\
			\object{NGC 3079} & $0.13 \pm 0.02$\tablefootmark{i} & $0.26 \pm 0.05$\tablefootmark{i} & $0.51 \pm 0.11$\tablefootmark{i} \\
			\object{NGC 3256} & $<0.065$ & $<0.20$ & $0.32 \pm 0.04$ \\
			\object{NGC 3690} & $<0.39$\tablefootmark{c} & ... & ... \\
			\object{NGC 4418} & $0.40 \pm 0.04$ & $0.65 \pm 0.09$ & $0.61 \pm 0.06$ \\
			\object{NGC 4945} & $0.09 \pm 0.01$ & $0.17 \pm 0.02$ & $0.51 \pm 0.02$ \\
			\object{NGC 5135} & $<0.26$ & ... & $<0.25$ \\
			\object{NGC 6946} & $<0.073$\tablefootmark{c} & $<0.10$\tablefootmark{c} & $0.71 \pm 0.1$ \\
			\object{NGC 7130} & $<0.18$ & $<0.25$ & $0.71 \pm 0.1$ \\
			\object{UGC 5101} & $<0.084$ & $<0.09 $ & $0.91 \pm 0.1$ \\
            \noalign{\smallskip}
            \hline

     \end{tabular}
     \tablefoot{
     The HC$_3$N line used is the 10\nobrkhyph9 line, whenever it is available. If another line is used, this is stated below. The errors are $1\sigma$ and have been calculated by error propagation from the errors given in Tables~\ref{tab:resultshc3n}\nobrkhyph\ref{tab:resultshnc}.\\
     \tablefoottext{a}{HC$_3$N value from \citet{aalto02} used.}
     \tablefoottext{b}{HC$_3$N value from this work used.}
     \tablefoottext{c}{HC$_3$N~12\nobrkhyph11 line.}
     \tablefoottext{d}{HC$_3$N~28\nobrkhyph27 line.}
     \tablefoottext{e}{HCN value from \citet{gracia08} used.}
     \tablefoottext{f}{HCN value from \citet{gao04} used.}
     \tablefoottext{g}{HNC value from \citet{perez07} used.}
     \tablefoottext{h}{HC$_3$N~16\nobrkhyph15 line.}
     \tablefoottext{i}{HCN and HNC values from \citet{perez07} used.}
	 }

\end{table}

The line ratios have been calculated using the method described in Appendix~\ref{app:nearby}, and are shown in Table~\ref{tab:ratios}. As already mentioned, the HC$_3$N/HCN and HC$_3$N/HNC line ratios of some of the most nearby galaxies will be somewhat overestimated due to their source size being larger than the beam size of the telescope. See Appendix~\ref{app:nearby2} for a discussion on this subject. The galaxies in our survey that do not fulfil the criterion $\theta_{\mathrm{s}} \lesssim \theta_{\mathrm{mb}}$ are \object{IC 342}, \object{M82}, \object{Maffei 2}, and \object{NGC 253}, and we therefore expect the real HC$_3$N/HCN and HC$_3$N/HNC ratios to be somewhat lower for these galaxies.

In Table~\ref{tab:ratios}, preference has been given to HC$_3$N~10\nobrkhyph9 lines before other HC$_3$N lines. Only if no 10\nobrkhyph9 line is available, another HC$_3$N line (specified in the footnotes) has been used for the ratios.

A few of the galaxies appear twice in Table~\ref{tab:ratios}. For these, several observations have been found for the same transition. As can be seen, the values of these observations do not always agree. If the HC$_3$N and HNC data are found in the same spectrum in one of the observations, preference has been given to this observation, as it will increase the accuracy on the HC$_3$N/HNC ratio. In all other cases, the spectrum of each observation has been investigated (when available), and the values from the spectra with the lowest noise levels have been given priority and are put first in Table~\ref{tab:ratios}.

\subsubsection{HC$_3$N-luminous galaxies}
\label{sec:hc3nlumpoor}

A definition of an HC$_3$N-luminous galaxy is now desirable. As HCN is the most common dense gas tracer, and also should be a more stable component of the dense gas than HNC, the HC$_3$N/HCN ratios were decided to be used for this definition. It seems like most galaxies have HC$_3$N/HCN ratios below 0.15, with the exception for a few interesting galaxies. Thus, we consider in the rest of the paper that galaxies with $\frac{I(\mathrm{HC}_3\mathrm{N})}{I(\mathrm{HCN})} > 0.15$ are HC$_3$N-luminous galaxies. If a corresponding limit should be set on the HC$_3$N/HNC ratio, it would be around 0.25 to include the same galaxies.

The galaxies thus seen as HC$_3$N-luminous or HC$_3$N-rich are \object{NGC 4418}, \object{IC 342}, \object{Circinus}, \object{M82}, \object{Maffei 2}, \object{Arp 220}, and \object{IRAS 17208-0014}. We also choose to include \object{IC 860}, considering that its moderate HC$_3$N/HNC ratio is for the HC$_3$N~28\nobrkhyph27 transition, as the higher transitions seem to be weaker than the $J=$10\nobrkhyph9 line in most galaxies where more than one line has been observed (when beam effects are compensated for). A few of these galaxies are quite nearby, and as discussed above, the HC$_3$N/HCN ratio of galaxies with source sizes larger than the telescope beam size will probably be overestimated. This is particularly the case for \object{M82}, \object{IC 342}, and \object{Maffei 2}.

Some galaxies can definitely be seen as HC$_3$N-poor, since they have HC$_3$N/HCN ratios (or upper limits for this ratio) less than or equal to 0.10: \object{NGC 253}, \object{NGC 1068}, \object{NGC 1808}, \object{NGC 3256}, \object{NGC 4945}, \object{NGC 6946}, and \object{UGC 5101}. Since \object{NGC 253} also belongs to the nearby galaxies, this value should probably be even lower.

\section{Discussion}
\label{sec:dis}

It is not surprising to find that most galaxies in our sample from published articles show HC$_3$N detections -- otherwise they would not be submitted for publication. However, if only counting the galaxies first investigated by us, there are only 6 HC$_3$N detections in the 19 galaxies. This sample of galaxies was nevertheless made to find a high number of HC$_3$N-luminous galaxies. This could mean one of three things:
\begin{itemize}
\item Our search-criteria for HC$_3$N-luminous galaxies are not appropriate.
\item The limit for HC$_3$N-luminous galaxies is set too high to include all ``interesting'' objects.
\item HC$_3$N-luminous galaxies are very rare, even among active galaxies.
\end{itemize}

\noindent
Several possible correlations between a high HC$_3$N intensity and other properties of the galaxies have been examined. This will be discussed below.

\subsection{In which types of galaxies do we find HC$_3$N?}

If the HC$_3$N-rich and -poor galaxies are compared with the galaxy classifications of Table~\ref{tab:galaxies}, the most obvious trend is that most of the HC$_3$N-poor galaxies are starbursts, with the exception for the LIRG \object{UGC 5101}. It is difficult to see any trend for the AGNs due to the low number of such objects in the sample.

When removing the nearby galaxies from the HC$_3$N-rich category due to their overestimated HC$_3$N/HCN-ratios mentioned earlier, the common denominator of the remaining galaxies seems to be that their source of activity is unknown or disputed -- they are labelled as ``obscured'' or ULIRGs. Thus, HC$_3$N might thrive in deeply obscured, shielded regions, where it cannot be destroyed by radiation. In starbursts, it might be destroyed by the strong UV field -- or not even created, as C$_2$H$_2$ on the grains will photo-dissociate into C$_2$H (see Section~\ref{sec:c2h2c2h}).

\subsection{Silicate absorption strength}
\label{sec:pah}

In \citet{spoon02}, several absorption features from ice and silicates as well as emission from PAHs in active galaxies are discussed. In \citet{spoon07}, an evolutionary plot for active galaxies is produced, showing two distinct regions in a plot over the equivalent width of the PAH 6.2~\hbox{\textmu}m emission line versus the strength of the silicate 9.7~\hbox{\textmu}m absorption band. Starburst galaxies tend to have a high PAH equivalent width, Seyfert galaxies have low PAH equivalent width and low silicate absorption strength, while ULIRGs have high silicate absorption strength and often also low PAH equivalent width.

By private communication with H.~W.~W.~Spoon, the numerical values for silicate absorption strength in all the galaxies in his sample were obtained. Most galaxies in our sample are also included in his sample. When comparing these values to our HC$_3$N/HCN ratios, a tentative pattern seems to appear.

\begin{figure}
    \centering
	\resizebox{\hsize}{!}{\includegraphics{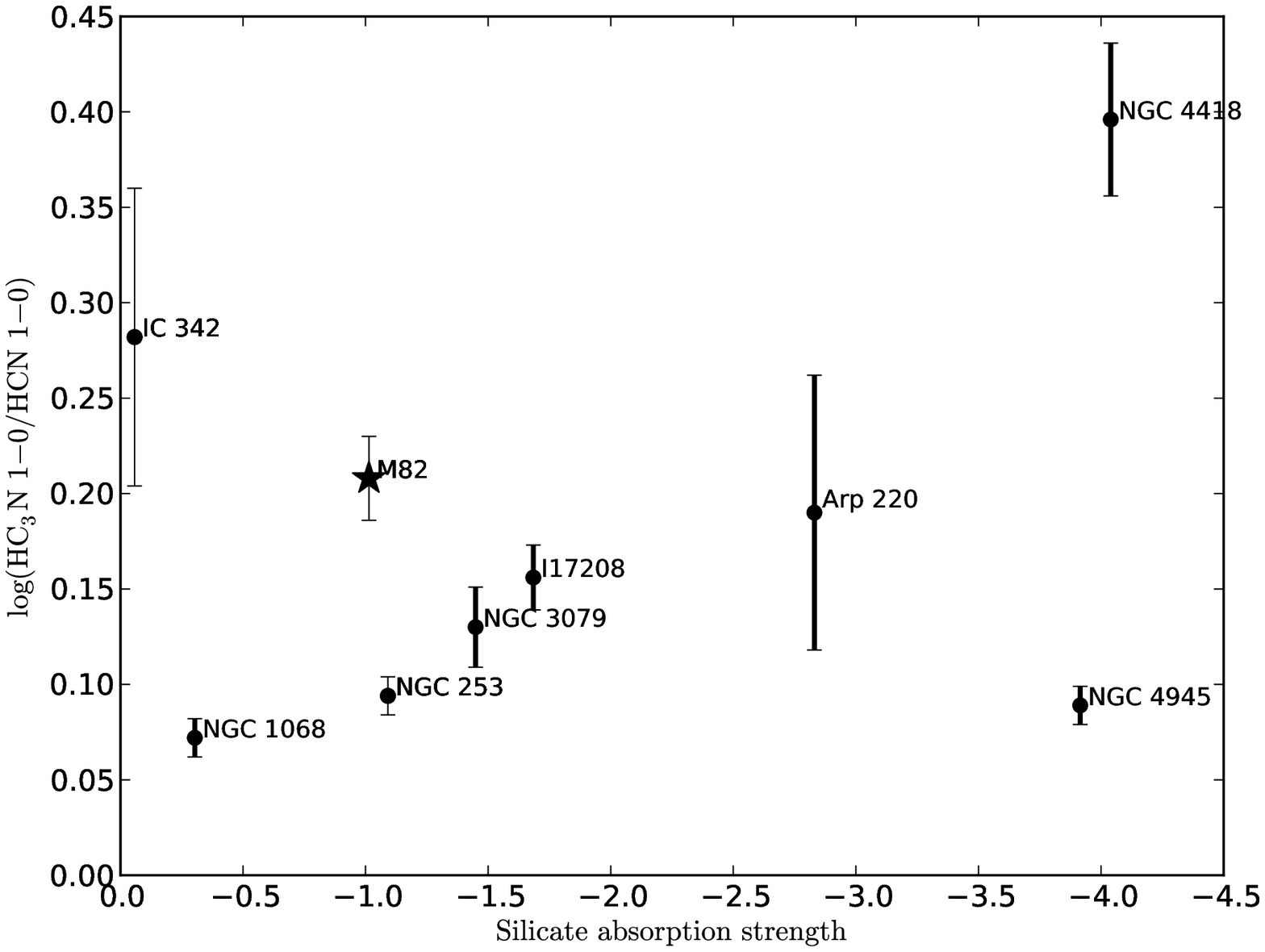}}
    \caption{Tentative correlation between HC$_3$N/HCN ratio and silicate absorption strength. The HC$_3$N/HCN values for \object{IC 342}, \object{NGC 253}, and \object{M82} are probably overestimated (indicated by thinner error bars, see Appendix~\ref{app:nearby2}). The objects for which the HC$_3$N~12\nobrkhyph11 line has been used instead of the HC$_3$N~10\nobrkhyph9 line when calculating the HC$_3$N/HCN ratio are indicated by a star. The silicate absorption strength is in a magnitude scale: A higher negative number means stronger silicate absorption.}
    \label{fig:hc3nsil}%
    \end{figure}

In Figure~\ref{fig:hc3nsil}, the relation between the HC$_3$N/HCN ratio and the silicate absorption strength is plotted. We note that two of the three nearby galaxies in the sample, \object{M82} and \object{IC 342}, are showing too high HC$_3$N/HCN ratios to fit into the pattern of the figure, which was expected (see Appendix~\ref{app:nearby2}). A correlation seems possible when excluding the nearby \object{M82}, \object{NGC 253}, and \object{IC 342} (correlation coefficient $r = -0.49$). One explanation to the correlation might be that HC$_3$N is formed in regions with silicates, where the silicates protect the HC$_3$N from radiation. Thus, HC$_3$N survives better in regions heavily obscured by silicates.

\subsection{Megamasers}
\label{sec:meg}

Neglecting the most nearby galaxies, where the HC$_3$N/HCN ratios probably are somewhat overestimated, all HC$_3$N-luminous galaxies have OH mega- or kilomasers \citep{darling06}. A few of the HC$_3$N-poor galaxies also have OH mega- or kilomaser activity (\object{NGC 253}, \object{NGC 1068}, and \object{UGC 5101}). The concept of defining megamaser strength only from its luminosity is however somewhat misleading. Firstly, a galaxy with much molecular gas is more likely to harbour a strong megamaser than one with little molecular gas -- the data in \citet{darling07} show a strong correlation between OH megamaser strength and CO and HCN luminosity. Thus, it is more reasonable to normalise the OH megamaser luminosity with some kind of luminosity for the molecular gas in the galaxy. The CO~1\nobrkhyph0 luminosity has been chosen for this, since CO is the primary tracer of molecular gas.

\begin{table}
\centering
      \caption[]{OH maser luminosities and CO luminosities for some of the sources in the sample.}
         \label{tab:ohmegamaser}
         \begin{tabular}{l c c c}
 
            \noalign{\smallskip}
			\hline
            \hline
            \noalign{\smallskip}
            Galaxy & $\log{L_{\mathrm{OH}}}$ & $L_{\mathrm{CO 1-0}}$ & $\frac{L_{\mathrm{OH}}}{L_{\mathrm{CO 1-0}}}$ \\
            & [$L_{\odot}$] & [$10^8$ K km s$^{-1}$ pc$^2$] & [$\frac{L_{\odot}}{10^8 \mathrm{K km s}^{-1} \mathrm{pc}^2}$] \\
            \noalign{\smallskip}
            \hline
            \noalign{\smallskip}
			\textbf{\object{Arp 220}} & 2.58 & 78.5 & 4.8 \\
			\textbf{\object{IC 342}} & None & ... & ... \\
			\object{IC 694}/\object{NGC 3690} & 1.38 & 29 & 0.83 \\
			\textbf{\object{IC 860}} & 0.27\tablefootmark{a} & 3.07\tablefootmark{b} & 0.61 \\
			\textbf{I17208} & 3.04  & 146.9 & 7.5 \\
			\textbf{\object{M82}} & -1.7 & 5.7 & 0.0035 \\
			\object{NGC 253} & -1.3 & 4.6 & 0.011 \\
			\object{NGC 1068} & -0.3 & 20.7 & 0.024 \\
			\object{NGC 1365} & None & ... & ... \\
			\object{NGC 1614} & None & ... & ... \\
			\object{NGC 2146} & None & ... & ... \\
			\object{NGC 3079} & Abs. & ... & ... \\
			\textbf{\object{NGC 4418}} & 0.04\tablefootmark{a} & 1.03\tablefootmark{c} & 1.1 \\
			\object{NGC 4945} & Abs. & ... & ... \\	
			\object{NGC 5135} & None & ... & ... \\
			\object{NGC 7130} & None & ... & ... \\		
			\object{UGC 5101} & 1.61 & 50.8 & 0.80 \\
		    \noalign{\smallskip}
            \hline

     \end{tabular}
	 
	 \tablefoot{
	 Data are from \citet{darling07} unless stated otherwise below. The note "None" in the $L_{OH}$ column indicates that \citet{darling07} has not detected any OH megamaser emission, and "Abs." that an OH absorber is detected. The names of the HC$_3$N-luminous galaxies are written in \textbf{boldface}. The HC$_3$N/HCN values for \object{IC 342} and \object{M82} are probably overestimated (see Appendix~\ref{app:nearby2}).
	 \tablefoottext{a}{From \citet{darling02}.}
	 \tablefoottext{b}{From observations by F. Costagliola.}
	 \tablefoottext{c}{From \citet{albrecht07}.}
	 }

\end{table}

    \begin{figure}
    \centering
	\resizebox{\hsize}{!}{\includegraphics{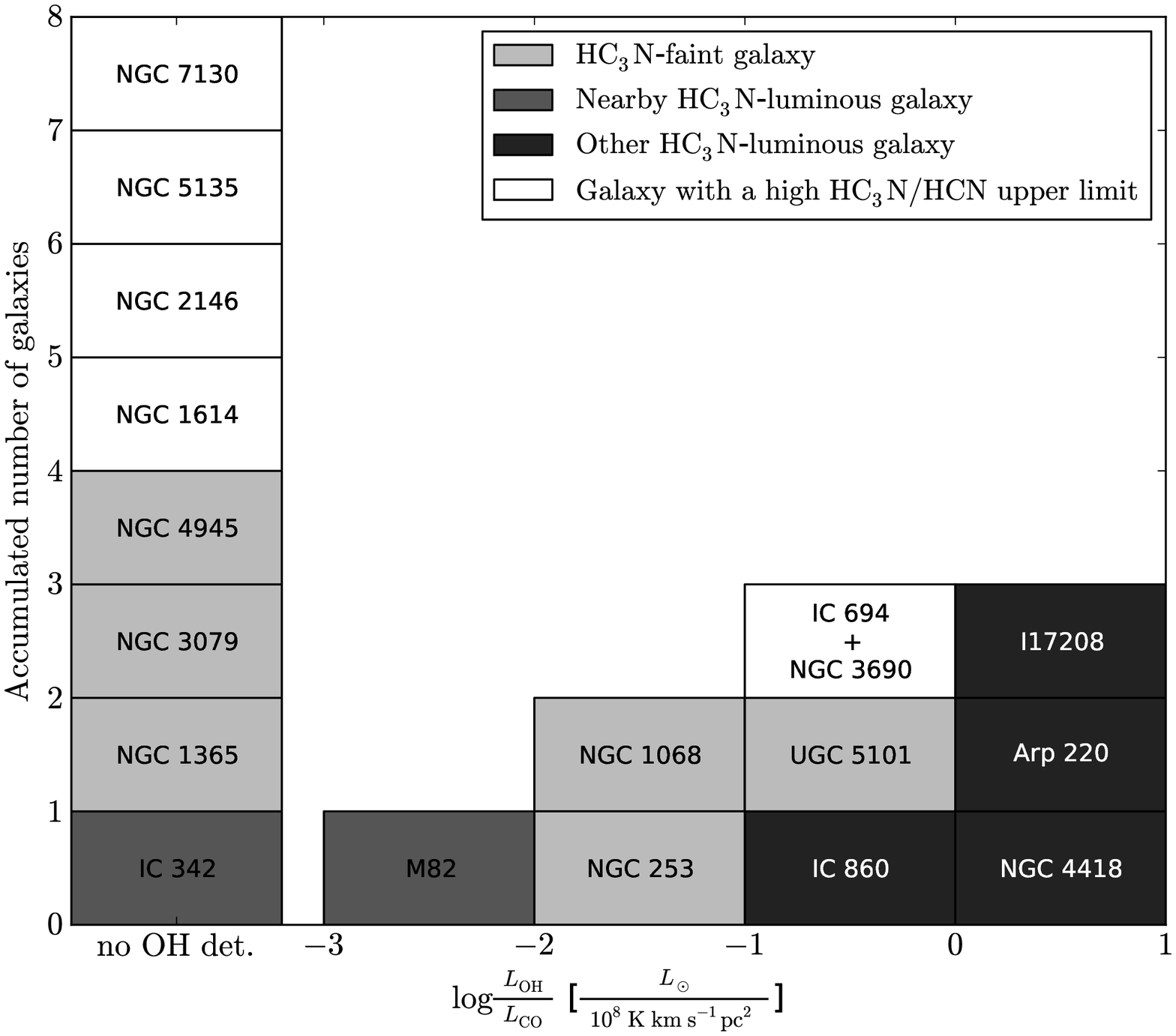}}
    \caption{Histogram over number of galaxies in normalised OH megamaser luminosity. Light-grey corresponds to galaxies considered HC$_3$N-poor, medium-gray corresponds to nearby HC$_3$N-luminous galaxies whose HC$_3$N/HCN ratios probably are overestimated (see Appendix~\ref{app:nearby2}), dark-grey corresponds to all other HC$_3$N-luminous galaxies, and white corresponds to galaxies where the upper limit on the HC$_3$N/HCN ratio is too high to tell whether it is HC$_3$N-luminous or not. The galaxies for which OH maser emission is non-detected in \citet{darling02} and/or \citet{darling07} are stacked in the leftmost column of the histogram.}
    \label{fig:ohbar}%
    \end{figure}

For \object{IC 860}, no CO~1\nobrkhyph0 luminosity value has been found in the literature, but from an observation performed with the IRAM~30~m telescope by F. Costagliola the luminosity could be calculated from the intensity with the method described in \citet{solomon97}. The used equation is:

\begin{equation}
L_{\mathrm{CO}} = 23.5 \Omega_{s\ast b} D_L^2 I_{\mathrm{CO}} (1 + z)^{-3}
\label{eq:solomon}
\end{equation}

\noindent
where $L_{\mathrm{CO}}$ is the line luminosity in K~km~s$^{-1}$~pc$^2$, $\Omega_{s\ast b}$ is the solid angle of the source convolved with the telescope beam in arcsec$^2$, $D_L$ is the luminosity distance in Mpc, $I_{\mathrm{CO}}$ is the main beam intensity of the line in K~km~s$^{-1}$, and $z$ is the redshift of the source. The intensity 9.83~K~km~s$^{-1}$, the beam width 22$''$, and the distance 59.1~Mpc gives a luminosity of $3.07 \cdot 10^8$~K~km~s$^{-1}$~pc$^2$, assuming that the source size is much smaller than the beam size, which is valid for \object{IC 860}.

In Table~\ref{tab:ohmegamaser}, the OH megamaser luminosity is compared with the CO luminosity. The data in this table is also shown in the form of a histogram in Figure~\ref{fig:ohbar}. The non-detections of OH megamasers and OH absorbers reported in \citet{darling07} are also listed, most of them being either HC$_3$N-poor or without any HC$_3$N detection (the exceptions are \object{IC~342} and \object{M82}, whose HC$_3$N/HCN ratios probably are overestimated, see Appendix~\ref{app:nearby2}).

It can be seen that the OH megamaser luminosities normalised with the galactic CO luminosities are much higher in the HC$_3$N-luminous galaxies than in the HC$_3$N-poor galaxies with an OH megamaser, especially when ignoring the nearby galaxies \object{M82} and \object{IC 342}. The average of the OH/CO luminosity ratios is more than 10 times higher in the HC$_3$N-luminous galaxies than in the HC$_3$N-poor galaxies. The non-detections (where HC$_3$N-poor galaxies are overrepresented) are not included in these averages. \object{UGC 5101} is the only HC$_3$N-poor galaxy with an OH megamaser that is strong compared to the amount of molecular gas in the galaxy. Also without the CO normalisation the trend can be seen clearly.
	
A possible explanation of the HC$_3$N correlating with OH megamasers is that the HC$_3$N is protected against destructive UV radiation by the warm dust which is needed to power the megamaser \citep{darling06}. Another possibility is that the HC$_3$N is pumped by the IR field caused by the warm dust, which also pumps the OH megamaser.

\subsection{IR flux density ratios}
\label{sec:iras}
\begin{figure}
    \centering
	\resizebox{\hsize}{!}{\includegraphics{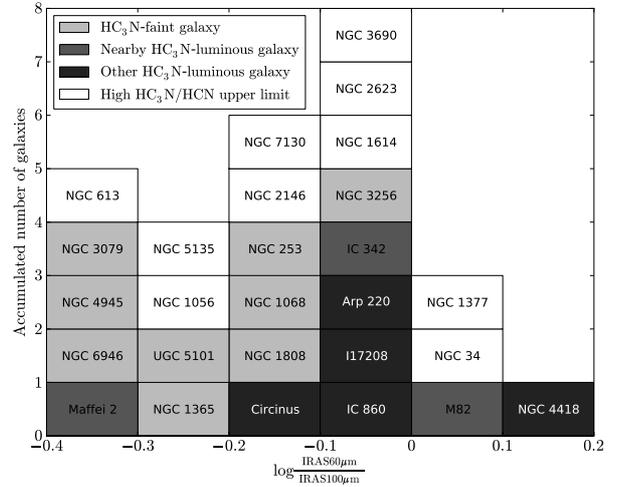}}
    \caption{Histogram over number of galaxies in IRAS 60~\hbox{\textmu}m/100~\hbox{\textmu}m flux density ratio. Light-grey corresponds to galaxies considered HC$_3$N-poor, medium-gray corresponds to nearby HC$_3$N-luminous galaxies whose HC$_3$N/HCN ratios probably are overestimated (see Appendix~\ref{app:nearby2}), dark-grey corresponds to all other HC$_3$N-luminous galaxies, and white corresponds to galaxies where the upper limit on the HC$_3$N/HCN ratio is too high to tell whether it is HC$_3$N-luminous or not. IRAS 60~\hbox{\textmu}m/100~\hbox{\textmu}m flux density values are all from \citet{sanders03}, except for \object{Circinus} and \object{Maffei 2}, which are from \citet{beichman88}.}
    \label{fig:iras}%
    \end{figure}

In several of the HC$_3$N rich galaxies, rotational-vibrational HC$_3$N lines have been detected \citep{costagliola10a,martin10,costagliola10b}. This suggests that IR pumping of the emission is present in these galaxies, which in turn indicates a warmer spectral energy distribution (SED) in these sources. We have thus compared the HC$_3$N/HCN ratios with the IRAS 60~\hbox{\textmu}m/100~\hbox{\textmu}m flux density ratios, but no linear correlation could be found. However, when plotting the data as a histogram (Figure~\ref{fig:iras}), we see a trend towards HC$_3$N-luminous galaxies having higher IRAS 60~\hbox{\textmu}m/100~\hbox{\textmu}m ratios, corresponding to warmer SEDs. \object{NGC 4418}, the object with the highest HC$_3$N/HCN ratio in the sample is also the galaxy with the highest IRAS 60~\hbox{\textmu}m/100~\hbox{\textmu}m ratio.

It should however be mentioned that the global 60~\hbox{\textmu}m/100~\hbox{\textmu}m flux density ratios might not be completely relevant, since they may tell more about the temperature on the extended dust distribution (and/or foreground dust) than the dust temperature in the nucleus, where the pumping likely occurs.

\subsection{\ion{C}{ii} flux}
\label{sec:cii}
\begin{figure}
    \centering
	\resizebox{\hsize}{!}{\includegraphics{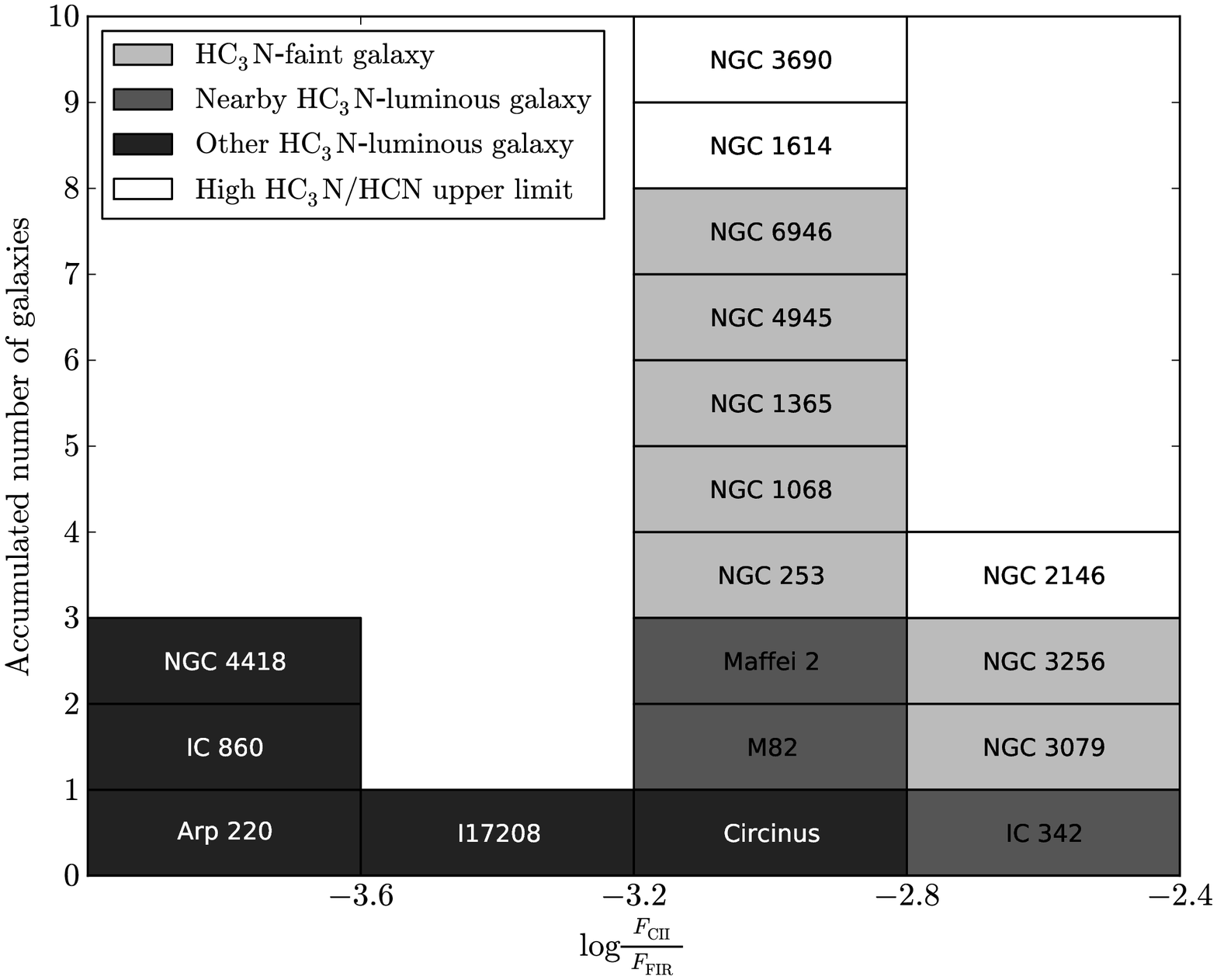}}
    \caption{Histogram over number of galaxies in \ion{C}{ii} flux normalised by FIR flux. Light-grey corresponds to galaxies considered HC$_3$N-poor, medium-gray corresponds to nearby HC$_3$N-luminous galaxies whose HC$_3$N/HCN ratios probably are overestimated (see Appendix~\ref{app:nearby2}), dark-grey corresponds to all other HC$_3$N-luminous galaxies, and white corresponds to galaxies where the upper limit on the HC$_3$N/HCN ratio is too high to tell whether it is HC$_3$N-luminous or not. FIR fluxes were calculated from IRAS 60{\textmu}m and 100{\textmu}m fluxes with the method described in \citet{bizyaev01}. The references for the used IRAS fluxes are given in the caption of Figure~\ref{fig:iras}. \ion{C}{ii} flux values are from \citet{negishi01}, except for \object{IC 342} and \object{NGC 3079} \citep{stacey91}; \object{IC 860} and \object{NGC 4418} \citep{malhotra01}; \object{Arp 220} \citep{luhman03}; and \object{NGC 1614} \citep{brauher09}.}
    \label{fig:cii}%
    \end{figure}

Many of the objects with high HC$_3$N/HCN ratios also have low \ion{C}{ii}/FIR flux ratios. This can be explained by that the C$^+$ ions are able to destroy HC$_3$N through reactions~\ref{eq:cii1} and \ref{eq:cii2} in Section~\ref{sec:destruct}. We have searched in the literature for available \ion{C}{ii} fluxes of the objects, which were found for more than half of the objects in our sample. With the limited amount of data, no linear correlation could be established. Instead, we display the data in the form of a histogram in Figure~\ref{fig:cii}, where we clearly see that a majority of the HC$_3$N-rich galaxies are very poor in \ion{C}{ii} flux. We also see that the nearby galaxies for which the high HC$_3$N/HCN ratio we expected to be an over-estimation due to beam effects (see Appendix~\ref{app:nearby2}) all belong to the \ion{C}{ii} rich part of the histogram. Also in this case, \object{NGC 4418} is the most extreme galaxy, with the lowest upper limit on the \ion{C}{ii}/FIR flux ratio.

\subsection{HNC/HCN}
\label{sec:hnc}

\begin{figure}
    \centering
	\resizebox{\hsize}{!}{\includegraphics{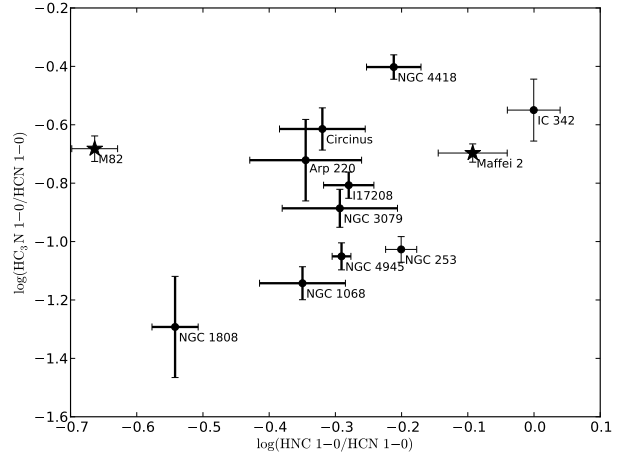}}
    \caption{log-log plot of the HC$_3$N/HCN line ratio versus the HNC/HCN line ratio. The HC$_3$N/HCN values for \object{IC 342}, \object{Maffei 2}, \object{M82}, and \object{NGC 253} are probably overestimated (indicated by thin error bars, see Appendix~\ref{app:nearby2}). The objects for which the HC$_3$N~12\nobrkhyph11 line has been used instead of the HC$_3$N~10\nobrkhyph9 line when calculating the HC$_3$N/HCN ratio are indicated by a star.}
    \label{fig:hc3nhcnhnc}%
\end{figure}

The HNC/HCN~1\nobrkhyph0 ratio is an indicator of the physical and chemical conditions in the dense molecular gas. Overluminous HNC is a sign of XDR chemistry \citep{aalto08}, while a low HNC/HCN ratio indicates shocks \citep{schilke92}. As is seen in Figure~\ref{fig:hc3nhcnhnc}, we find a correlation between the HC$_3$N/HCN and HNC/HCN~1\nobrkhyph0 line ratios. The HC$_3$N/HCN ratios of \object{M82}, \object{Maffei 2}, \object{IC 342}, and \object{NGC 253} are probably overestimated (see Appendix~\ref{app:nearby2}), and the correlation coefficient is $r = 0.65$ when these four objects are excluded. This correlation will be further discussed in \citet{costagliola10b}.

Attempts were also made trying to find a correlation between the HC$_3$N ratios and ratios of higher HNC and HCN transitions, e.g. the HNC/HCN~3\nobrkhyph2 ratio. However, too few HNC~3\nobrkhyph2 and HCN~3\nobrkhyph2 spectra for the galaxies in the sample are available in the literature -- with only five data points no conclusions can be drawn. It was although noticed that the HCN~3\nobrkhyph2 and HNC~3\nobrkhyph2 intensities seem to be very uncertain, at least for \object{Arp 220} and \object{NGC 4418}. Shortly, it seems like the different instruments used for the observations affect the measured value to a non-negligible extent. This is further discussed in \citet{lindberg09}.

We were not able to reproduce the weak correlation between the HNC/HCN~1\nobrkhyph0 ratio and the FIR luminosity described in \cite{aalto02}.

\subsection{Future observational tests}
\label{sec:fut}

We have found a strong trend between the HC$_3$N/HCN ratio and OH megamaser activity (see Section~\ref{sec:meg}). By studying different excitation levels of HC$_3$N in these and other OH megamaser galaxies, the cause of this correlation can be investigated. If HC$_3$N is pumped, higher transitions, including vibrational transitions, should be found. This has already been detected in \object{NGC 4418} \citep{costagliola10a}.

To better establish the importance of HC$_3$N as an indicator of activity in certain galaxies, it is important to do further line surveys of HC$_3$N, HCN, and HNC, as well as other molecules who trace the properties of the molecular gas, such as HCO$^+$ and C$_2$H. Such a line survey could also test our weak correlation between the HC$_3$N/HCN and HNC/HCN line ratios, and perhaps finding other correlations which would enable a better understanding of the chemistry in obscured and active galaxies. A new line survey was made with the new EMIR receiver at the IRAM~30~m telescope in June 2009, and the results from this survey will be published in \citet{costagliola10b}. The sample in this survey has been chosen to get more HC$_3$N data on sources in Spoon's sample to test the possible correlations between the HC$_3$N/HCN ratio and PAH equivalent width and/or silicate absorption strength. The bandwidth of the EMIR receiver allows for several spectral lines being observed in the same spectrum, and thus many different molecular species can be observed at the same time.

It might also be interesting to search for other long-carbon-chain molecules in the HC$_3$N-luminous galaxies, e.g. HC$_5$N, C$_2$H, C$_4$H, C$_3$H$_2$, and C$_4$H$_2$, which have all been found in star-forming regions in the Galaxy \citep{sakai08,sakai09}.

Finally, we suggest mapping of HC$_3$N, C$_2$H, HCN, and HNC in a larger number of galaxies, especially in the HC$_3$N-luminous galaxies, to compare the results with the maps of \object{IC 342} in \citet{meier05}. The HC$_3$N absorption lines detected in a $z\sim0.89$ galaxy in front of PKS~1830\nobrkhyph211 \citep{henkel09} indicate that HC$_3$N is not only present in the core of a galaxy, but may also be present in the disc.

\section{Conclusions}
\label{sec:con}
We have presented the first survey of HC$_3$N observations in extragalactic objects. The main conclusions from this survey are as follows:
   \begin{enumerate}
		\item Bright HC$_3$N emission is rather uncommon in galaxies. It was only detected in 6 of the 19 galaxies which had not been investigated before, even though that sample was selected to find many HC$_3$N-luminous galaxies.
		\item Most HC$_3$N-luminous galaxies are obscured galaxies. Starburst galaxies seem to be poor in HC$_3$N. There are too few AGN galaxies in the sample to tell if these normally are rich or poor in HC$_3$N.
      	\item Weak correlations can be seen between the HC$_3$N/HCN ratio and silicate 9.7~\hbox{\textmu}m absorption strength.
      	\item There is a strong correlation between OH megamaser activity and HC$_3$N luminosity. Most HC$_3$N-luminous galaxies have an OH megamaser. This could be related to a high dust obscuration in the HC$_3$N-luminous galaxies.
       	\item There is a connection between the HC$_3$N/HCN ratio and the IRAS 60~\hbox{\textmu}m/100~\hbox{\textmu}m flux density ratios, indicating a higher dust temperature in these galaxies, which could cause vibrational excitation of the HC$_3$N molecule.
       	\item There is a strong connection between a high HC$_3$N/HCN ratio and a low \ion{C}{ii}/FIR flux ratio in the studied objects. This could be explained by C$^+$ ions being required to destroy the HC$_3$N molecule. 
		\item There is a correlation between the HC$_3$N/HCN and HNC/HCN line ratios.
   \end{enumerate}

\begin{acknowledgements}
		Many thanks to the IRAM, OSO, and SEST staff for their help during the observations. We are grateful to H.~W.~W.~Spoon for sharing his PAH and silicate data with us. We would also like to thank the anonymous referee for several useful suggestions which improved the manuscript.
\end{acknowledgements}

   \begin{figure*}
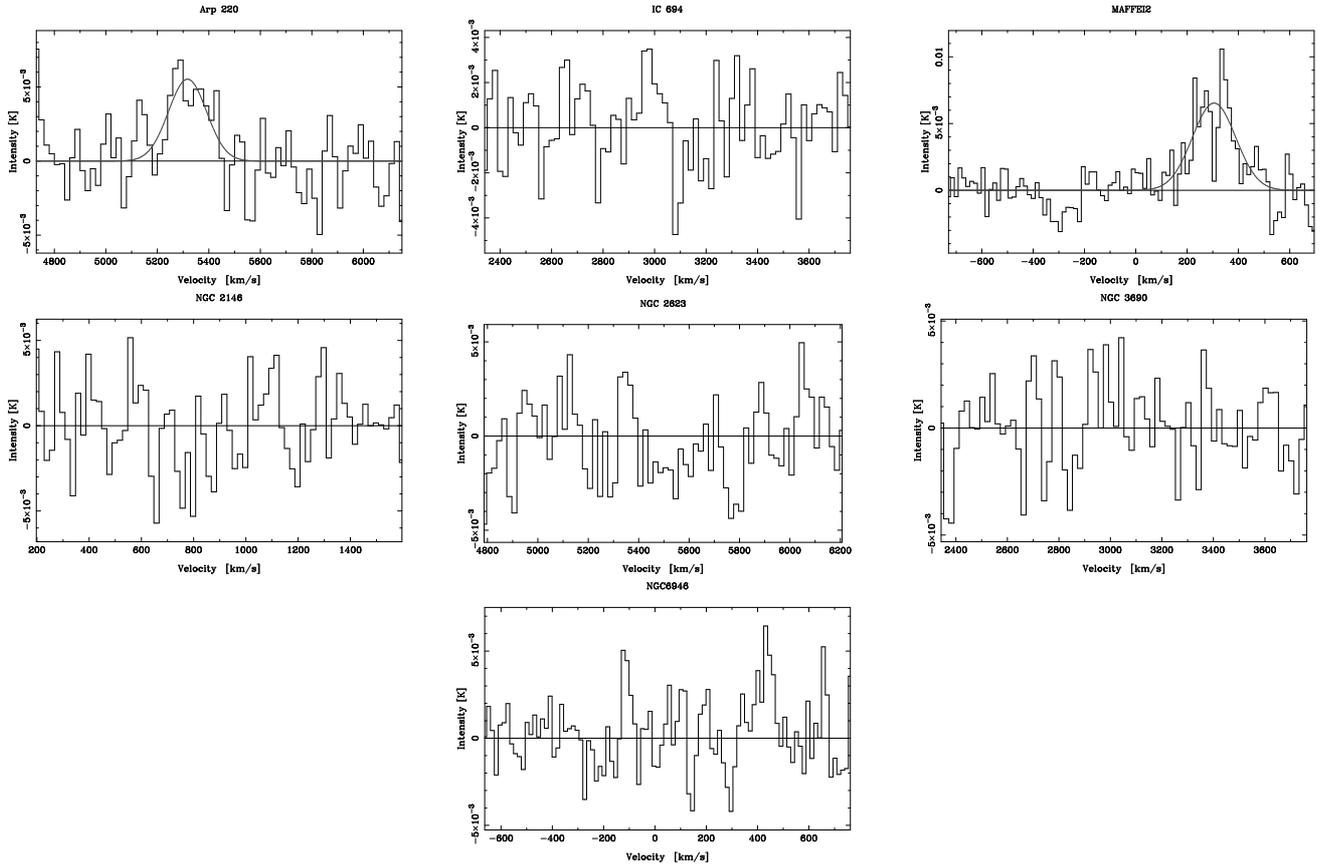

   \centering
   $\begin{array}{c@{\hspace{0.7cm}}c@{\hspace{0.7cm}}c}
	\includegraphics[width=5.2cm]{15565fg7a.eps}  &
	\includegraphics[width=5.2cm]{15565fg7b.eps} &
	\hspace{0.2cm}
	\includegraphics[width=5.2cm]{15565fg7c.eps} \\
	\includegraphics[width=5.2cm]{15565fg7d.eps}  &
	\includegraphics[width=5.2cm]{15565fg7e.eps} &
	\includegraphics[width=5.2cm]{15565fg7f.eps} \\
	\hspace{0.15cm}
	&
	\includegraphics[width=5.2cm]{15565fg7g.eps} \\
	\end{array}$
   \caption{HC$_3$N~12\nobrkhyph11 spectra for \object{Arp 220} (detection), \object{IC 694} (non-detection), \object{Maffei 2} (detection), \object{NGC 2146}, \object{NGC 2623}, \object{NGC 3690}, and \object{NGC 6946} (non-detections).}
              \label{fig:hc3n12-11}%
    \end{figure*}

   \begin{figure*}
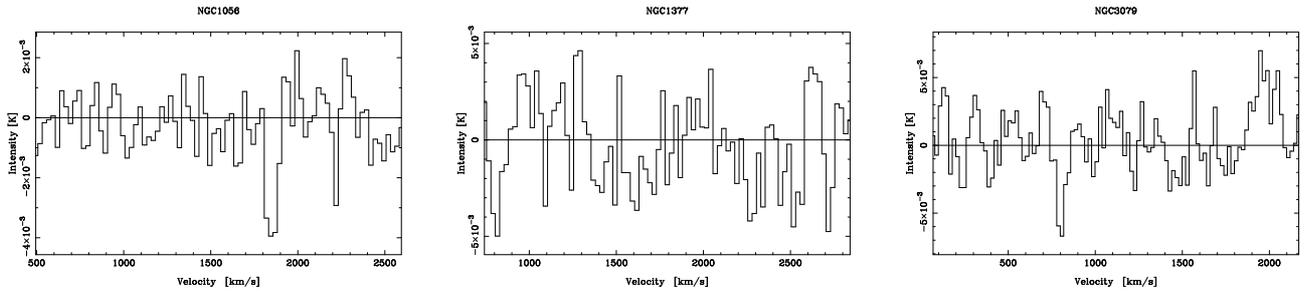

   \centering
   $\begin{array}{c@{\hspace{0.7cm}}c@{\hspace{0.7cm}}c}
	\includegraphics[width=5.2cm]{15565fg8a.eps}  &  
	\includegraphics[width=5.2cm]{15565fg8b.eps}  &
	\includegraphics[width=5.2cm]{15565fg8c.eps} \\
	\end{array}$
   \caption{HC$_3$N~16\nobrkhyph15 spectra for \object{NGC 1056}, \object{NGC 1377}, and \object{NGC 3079} (non-detections). The absorption-like features in the \object{NGC 1056} spectrum are most likely due to a problem with the backend of the receiver.}
              \label{fig:hc3n16-15}%
    \end{figure*}

   \begin{figure*}
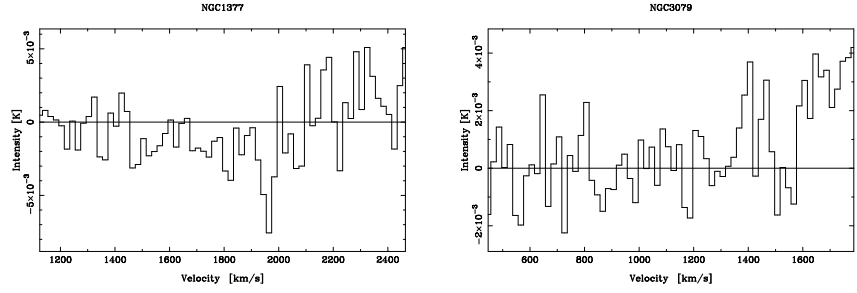

   \centering
   $\begin{array}{c@{\hspace{0.7cm}}c@{\hspace{0.7cm}}c}
	\hspace{0.6cm}	
	\includegraphics[width=5.2cm]{15565fg9a.eps} &	
	\includegraphics[width=5.2cm]{15565fg9b.eps} \\
	\end{array}$
   \caption{HC$_3$N~25\nobrkhyph24 spectra for \object{NGC 1377} and \object{NGC 3079} (non-detections).}
              \label{fig:hc3n25-24}%
    \end{figure*}
 
   \begin{figure*}
   \centering
   $\begin{array}{c@{\hspace{0.7cm}}c@{\hspace{0.7cm}}c}
   	&
	\includegraphics[width=5.2cm]{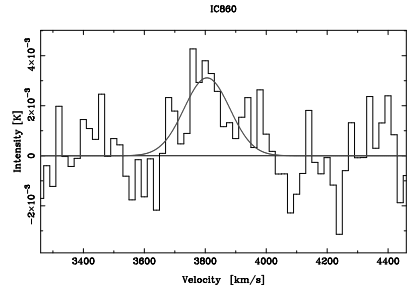} \\
	\end{array}$
   \caption{HC$_3$N~28\nobrkhyph27 spectrum for \object{IC 860} (detection).}
              \label{fig:hc3n28-27}%
    \end{figure*}

   \begin{figure*}
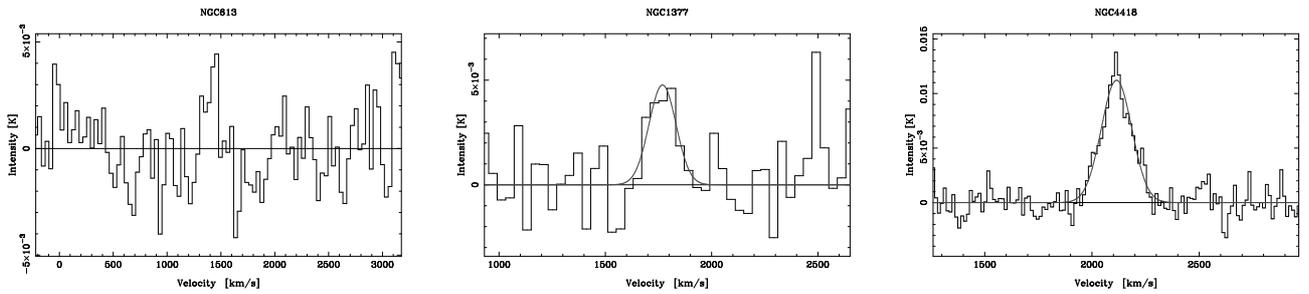

   \centering
   $\begin{array}{c@{\hspace{0.7cm}}c@{\hspace{0.7cm}}c}
	\includegraphics[width=5.2cm]{15565fg11a.eps}  &  
	\includegraphics[width=5.2cm]{15565fg11b.eps}  &
	\includegraphics[width=5.2cm]{15565fg11c.eps} \\
	\end{array}$
   \caption{HCN~1\nobrkhyph0 spectra for \object{NGC 613} (non-detection), \object{NGC 1377}, and \object{NGC 4418} (detections).}
              \label{fig:hcn1-0}%
    \end{figure*}
    
\bibliographystyle{aa} 
\bibliography{15565}

\begin{thebibliography}{73}
\expandafter\ifx\csname natexlab\endcsname\relax\def\natexlab#1{#1}\fi

\bibitem[{{Aalto}(2008)}]{aalto08}
{Aalto}, S. 2008, \apss, 313, 273

\bibitem[{{Aalto} {et~al.}(1995){Aalto}, {Booth}, {Black}, \&
  {Johansson}}]{aalto95}
{Aalto}, S., {Booth}, R.~S., {Black}, J.~H., \& {Johansson}, L.~E.~B. 1995,
  \aap, 300, 369

\bibitem[{{Aalto} {et~al.}(1994){Aalto}, {Booth}, {Black}, {Koribalski}, \&
  {Wielebinski}}]{aalto94}
{Aalto}, S., {Booth}, R.~S., {Black}, J.~H., {Koribalski}, B., \&
  {Wielebinski}, R. 1994, \aap, 286, 365

\bibitem[{{Aalto} {et~al.}(2007){Aalto}, {Monje}, \&
  {Mart{\'{\i}}n}}]{aalto07a}
{Aalto}, S., {Monje}, R., \& {Mart{\'{\i}}n}, S. 2007, \aap, 475, 479

\bibitem[{{Aalto} {et~al.}(2002){Aalto}, {Polatidis}, {H{\"u}ttemeister}, \&
  {Curran}}]{aalto02}
{Aalto}, S., {Polatidis}, A.~G., {H{\"u}ttemeister}, S., \& {Curran}, S.~J.
  2002, \aap, 381, 783

\bibitem[{{Aalto} {et~al.}(1997){Aalto}, {Radford}, {Scoville}, \&
  {Sargent}}]{aalto97}
{Aalto}, S., {Radford}, S.~J.~E., {Scoville}, N.~Z., \& {Sargent}, A.~I. 1997,
  \apjl, 475, L107

\bibitem[{{Albrecht} {et~al.}(2007){Albrecht}, {Kr{\"u}gel}, \&
  {Chini}}]{albrecht07}
{Albrecht}, M., {Kr{\"u}gel}, E., \& {Chini}, R. 2007, \aap, 462, 575

\bibitem[{{Baan} {et~al.}(2008){Baan}, {Henkel}, {Loenen}, {Baudry}, \&
  {Wiklind}}]{baan08}
{Baan}, W.~A., {Henkel}, C., {Loenen}, A.~F., {Baudry}, A., \& {Wiklind}, T.
  2008, \aap, 477, 747

\bibitem[{{Bajaja} {et~al.}(1995){Bajaja}, {Wielebinski}, {Reuter}, {Harnett},
  \& {Hummel}}]{bajaja95}
{Bajaja}, E., {Wielebinski}, R., {Reuter}, H.-P., {Harnett}, J.~I., \&
  {Hummel}, E. 1995, \aaps, 114, 147

\bibitem[{{Beichman} {et~al.}(1988){Beichman}, {Neugebauer}, {Habing}, {Clegg},
  \& {Chester}}]{beichman88}
{Beichman}, C.~A., {Neugebauer}, G., {Habing}, H.~J., {Clegg}, P.~E., \&
  {Chester}, T.~J., eds. 1988, {Infrared astronomical satellite (IRAS) catalogs
  and atlases. Volume 1: Explanatory supplement}, Vol.~1

\bibitem[{{Bizyaev}(2001)}]{bizyaev01}
{Bizyaev}, D. 2001, \apss, 276, 775

\bibitem[{{Bohme} \& {Raksit}(1985)}]{bohme85}
{Bohme}, D.~K. \& {Raksit}, A.~B. 1985, \mnras, 213, 717

\bibitem[{{Brauher} {et~al.}(2009){Brauher}, {Dale}, \& {Helou}}]{brauher09}
{Brauher}, J.~R., {Dale}, D.~A., \& {Helou}, G. 2009, VizieR Online Data
  Catalog, 217, 80280

\bibitem[{{Bryant} \& {Scoville}(1999)}]{bryant99}
{Bryant}, P.~M. \& {Scoville}, N.~Z. 1999, \aj, 117, 2632

\bibitem[{{Casoli} {et~al.}(1992){Casoli}, {Dupraz}, \& {Combes}}]{casoli92}
{Casoli}, F., {Dupraz}, C., \& {Combes}, F. 1992, \aap, 264, 49

\bibitem[{{Chapman} {et~al.}(2009){Chapman}, {Millar}, {Wardle}, {Burton}, \&
  {Walsh}}]{chapman09}
{Chapman}, J.~F., {Millar}, T.~J., {Wardle}, M., {Burton}, M.~G., \& {Walsh},
  A.~J. 2009, \mnras, 394, 221

\bibitem[{{Cherchneff} {et~al.}(1993){Cherchneff}, {Glassgold}, \&
  {Mamon}}]{cherchneff93}
{Cherchneff}, I., {Glassgold}, A.~E., \& {Mamon}, G.~A. 1993, \apj, 410, 188

\bibitem[{{Costagliola} \& {Aalto}(2010a)}]{costagliola10a}
{Costagliola}, F. \& {Aalto}, S. 2010a, \aap, 515, A71

\bibitem[{{Costagliola} {et~al.}(2010b){Costagliola}, {Aalto}, {Rodriguez},
  {Spoon}, {Mart{\'{\i}}n}, {Perez-Torres}, {Alberdi}, {Lindberg}, {Batejat},
  {J\"{u}tte}, \& {Lahuis}}]{costagliola10b}
{Costagliola}, F., {Aalto}, S., {Rodriguez}, M.~I., {et~al.} 2010b, {submitted
  to \aap}

\bibitem[{{Curran} {et~al.}(2000){Curran}, {Aalto}, \& {Booth}}]{curran00}
{Curran}, S.~J., {Aalto}, S., \& {Booth}, R.~S. 2000, \aaps, 141, 193

\bibitem[{{Curran} {et~al.}(2001a){Curran}, {Johansson}, {Bergman},
  {Heikkil{\"a}}, \& {Aalto}}]{curran01a}
{Curran}, S.~J., {Johansson}, L.~E.~B., {Bergman}, P., {Heikkil{\"a}}, A., \&
  {Aalto}, S. 2001a, \aap, 367, 457

\bibitem[{{Curran} {et~al.}(2001b){Curran}, {Polatidis}, {Aalto}, \&
  {Booth}}]{curran01b}
{Curran}, S.~J., {Polatidis}, A.~G., {Aalto}, S., \& {Booth}, R.~S. 2001b,
  \aap, 373, 459

\bibitem[{{Dale} {et~al.}(2005){Dale}, {Sheth}, {Helou}, {Regan}, \&
  {H{\"u}ttemeister}}]{dale05}
{Dale}, D.~A., {Sheth}, K., {Helou}, G., {Regan}, M.~W., \& {H{\"u}ttemeister},
  S. 2005, \aj, 129, 2197

\bibitem[{{Darling}(2007)}]{darling07}
{Darling}, J. 2007, \apjl, 669, L9

\bibitem[{{Darling} \& {Giovanelli}(2002)}]{darling02}
{Darling}, J. \& {Giovanelli}, R. 2002, \aj, 124, 100

\bibitem[{{Darling} \& {Giovanelli}(2006)}]{darling06}
{Darling}, J. \& {Giovanelli}, R. 2006, \aj, 132, 2596

\bibitem[{{de~Vicente} {et~al.}(2000){de~Vicente}, {Mart{\'{\i}}n-Pintado},
  {Neri}, \& {Colom}}]{devicente00}
{de~Vicente}, P., {Mart{\'{\i}}n-Pintado}, J., {Neri}, R., \& {Colom}, P. 2000,
  \aap, 361, 1058

\bibitem[{{Fukuzawa} \& {Osamura}(1997)}]{fukuzawa97}
{Fukuzawa}, K. \& {Osamura}, Y. 1997, \apj, 489, 113

\bibitem[{{Gao} \& {Solomon}(2004)}]{gao04}
{Gao}, Y. \& {Solomon}, P.~M. 2004, \apjs, 152, 63

\bibitem[{{Graci{\'a}-Carpio} {et~al.}(2006){Graci{\'a}-Carpio},
  {Garc{\'{\i}}a-Burillo}, {Planesas}, \& {Colina}}]{gracia06}
{Graci{\'a}-Carpio}, J., {Garc{\'{\i}}a-Burillo}, S., {Planesas}, P., \&
  {Colina}, L. 2006, \apjl, 640, L135

\bibitem[{{Graci{\'a}-Carpio} {et~al.}(2008){Graci{\'a}-Carpio},
  {Garc{\'{\i}}a-Burillo}, {Planesas}, {Fuente}, \& {Usero}}]{gracia08}
{Graci{\'a}-Carpio}, J., {Garc{\'{\i}}a-Burillo}, S., {Planesas}, P., {Fuente},
  A., \& {Usero}, A. 2008, \aap, 479, 703

\bibitem[{{Heikkil{\"a}} {et~al.}(1999){Heikkil{\"a}}, {Johansson}, \&
  {Olofsson}}]{heikkila99}
{Heikkil{\"a}}, A., {Johansson}, L.~E.~B., \& {Olofsson}, H. 1999, \aap, 344,
  817

\bibitem[{{Henkel} {et~al.}(2009){Henkel}, {Menten}, {Murphy}, {Jethava},
  {Flambaum}, {Braatz}, {Muller}, {Ott}, \& {Mao}}]{henkel09}
{Henkel}, C., {Menten}, K.~M., {Murphy}, M.~T., {et~al.} 2009, \aap, 500, 725

\bibitem[{{Henkel} {et~al.}(1988){Henkel}, {Schilke}, \&
  {Mauersberger}}]{henkel88}
{Henkel}, C., {Schilke}, P., \& {Mauersberger}, R. 1988, \aap, 201, L23

\bibitem[{{H\"{u}ttemeister} {et~al.}(1995){H\"{u}ttemeister}, {Henkel},
  {Mauersberger}, {Brouillet}, {Wiklind}, \& {Millar}}]{huttemeister95}
{H\"{u}ttemeister}, S., {Henkel}, C., {Mauersberger}, R., {et~al.} 1995, \aap,
  295, 571

\bibitem[{{Imanishi} {et~al.}(2004){Imanishi}, {Nakanishi}, {Kuno}, \&
  {Kohno}}]{imanishi04}
{Imanishi}, M., {Nakanishi}, K., {Kuno}, N., \& {Kohno}, K. 2004, \aj, 128,
  2037

\bibitem[{{Irvine} {et~al.}(1987){Irvine}, {Goldsmith}, \&
  {Hjalmarson}}]{irvine87}
{Irvine}, W.~M., {Goldsmith}, P.~F., \& {Hjalmarson}, {\AA}. 1987, in
  Astrophysics and Space Science Library, Vol. 134, Interstellar Processes, ed.
  D.~J. {Hollenbach} \& H.~A. {Thronson}, Jr., 561

\bibitem[{{Knudsen} {et~al.}(2007){Knudsen}, {Walter}, {Weiss}, {Bolatto},
  {Riechers}, \& {Menten}}]{knudsen07}
{Knudsen}, K.~K., {Walter}, F., {Weiss}, A., {et~al.} 2007, \apj, 666, 156

\bibitem[{{Koda} {et~al.}(2002){Koda}, {Sofue}, {Kohno}, {Nakanishi},
  {Onodera}, {Okumura}, \& {Irwin}}]{koda02}
{Koda}, J., {Sofue}, Y., {Kohno}, K., {et~al.} 2002, \apj, 573, 105

\bibitem[{{Kohno} {et~al.}(2001){Kohno}, {Matsushita}, {Vila-Vilar{\'o}},
  {Okumura}, {Shibatsuka}, {Okiura}, {Ishizuki}, \& {Kawabe}}]{kohno01}
{Kohno}, K., {Matsushita}, S., {Vila-Vilar{\'o}}, B., {et~al.} 2001, in
  Astronomical Society of the Pacific Conference Series, Vol. 249, The Central
  Kiloparsec of Starbursts and AGN: The La Palma Connection, ed. {J.~H.~Knapen,
  J.~E.~Beckman, I.~Shlosman, \& T.~J.~Mahoney}, 672

\bibitem[{{Krips} {et~al.}(2008){Krips}, {Neri}, {Garc{\'{\i}}a-Burillo},
  {Mart{\'{\i}}n}, {Combes}, {Graci{\'a}-Carpio}, \& {Eckart}}]{krips08}
{Krips}, M., {Neri}, R., {Garc{\'{\i}}a-Burillo}, S., {et~al.} 2008, \apj, 677,
  262

\bibitem[{{Lepp} \& {Dalgarno}(1996)}]{lepp96}
{Lepp}, S. \& {Dalgarno}, A. 1996, \aap, 306, L21

\bibitem[{{Lindberg}(2009)}]{lindberg09}
{Lindberg}, J. 2009, Master's thesis, Chalmers~Univ.~Technol., G\"{o}teborg

\bibitem[{{Luhman} {et~al.}(2003){Luhman}, {Satyapal}, {Fischer}, {Wolfire},
  {Sturm}, {Dudley}, {Lutz}, \& {Genzel}}]{luhman03}
{Luhman}, M.~L., {Satyapal}, S., {Fischer}, J., {et~al.} 2003, \apj, 594, 758

\bibitem[{{Malhotra} {et~al.}(2001){Malhotra}, {Kaufman}, {Hollenbach},
  {Helou}, {Rubin}, {Brauher}, {Dale}, {Lu}, {Lord}, {Stacey}, {Contursi},
  {Hunter}, \& {Dinerstein}}]{malhotra01}
{Malhotra}, S., {Kaufman}, M.~J., {Hollenbach}, D., {et~al.} 2001, \apj, 561,
  766

\bibitem[{{Maloney} {et~al.}(1996){Maloney}, {Hollenbach}, \&
  {Tielens}}]{maloney96}
{Maloney}, P.~R., {Hollenbach}, D.~J., \& {Tielens}, A.~G.~G.~M. 1996, \apj,
  466, 561

\bibitem[{{Mart{\'{\i}}n} {et~al.}(2010){Mart{\'{\i}}n}, {Krips},
  {Mart{\'{\i}}n-Pintado}, {Aalto}, {Zhao}, {Peck}, {Petitpas}, {Monje}, \&
  {An}}]{martin10}
{Mart{\'{\i}}n}, S., {Krips}, M., {Mart{\'{\i}}n-Pintado}, J., {et~al.} 2010,
  {submitted to \aap}

\bibitem[{{Mart{\'{\i}}n} {et~al.}(2006){Mart{\'{\i}}n}, {Mauersberger},
  {Mart{\'{\i}}n-Pintado}, {Henkel}, \& {Garc{\'{\i}}a-Burillo}}]{martin06}
{Mart{\'{\i}}n}, S., {Mauersberger}, R., {Mart{\'{\i}}n-Pintado}, J., {Henkel},
  C., \& {Garc{\'{\i}}a-Burillo}, S. 2006, \apjs, 164, 450

\bibitem[{{Mauersberger} {et~al.}(1990){Mauersberger}, {Henkel}, \&
  {Sage}}]{mauersberger90}
{Mauersberger}, R., {Henkel}, C., \& {Sage}, L.~J. 1990, \aap, 236, 63

\bibitem[{{Meier} \& {Turner}(2005)}]{meier05}
{Meier}, D.~S. \& {Turner}, J.~L. 2005, \apj, 618, 259

\bibitem[{{Meijerink} \& {Spaans}(2005)}]{meijerink05}
{Meijerink}, R. \& {Spaans}, M. 2005, \aap, 436, 397

\bibitem[{{Meijerink} {et~al.}(2007){Meijerink}, {Spaans}, \&
  {Israel}}]{meijerink07}
{Meijerink}, R., {Spaans}, M., \& {Israel}, F.~P. 2007, \aap, 461, 793

\bibitem[{{NED}(2009)}]{ned}
{NED}. 2009, NASA Extragalactic Database

\bibitem[{{Negishi} {et~al.}(2001){Negishi}, {Onaka}, {Chan}, \&
  {Roellig}}]{negishi01}
{Negishi}, T., {Onaka}, T., {Chan}, K., \& {Roellig}, T.~L. 2001, \aap, 375,
  566

\bibitem[{{Nguyen-Q-Rieu} {et~al.}(1989){Nguyen-Q-Rieu}, {Nakai}, \&
  {Jackson}}]{nguyen89}
{Nguyen-Q-Rieu}, {Nakai}, N., \& {Jackson}, J.~M. 1989, \aap, 220, 57

\bibitem[{{Nguyen-Rieu} {et~al.}(1994){Nguyen-Rieu}, {Viallefond}, {Combes},
  {Jackson}, {Lequeux}, {Radford}, \& {Truong-Bach}}]{nguyen94}
{Nguyen-Rieu}, {Viallefond}, F., {Combes}, F., {et~al.} 1994, in Astronomical
  Society of the Pacific Conference Series, Vol.~59, IAU Colloq. 140: Astronomy
  with Millimeter and Submillimeter Wave Interferometry, ed. M.~{Ishiguro} \&
  J.~{Welch}, 336

\bibitem[{{Papadopoulos}(2007)}]{papadopoulos07}
{Papadopoulos}, P.~P. 2007, \apj, 656, 792

\bibitem[{{P{\'e}rez-Beaupuits} {et~al.}(2007){P{\'e}rez-Beaupuits}, {Aalto},
  \& {Gerebro}}]{perez07}
{P{\'e}rez-Beaupuits}, J.~P., {Aalto}, S., \& {Gerebro}, H. 2007, \aap, 476,
  177

\bibitem[{{Prasad} \& {Huntress}(1980)}]{prasad80}
{Prasad}, S.~S. \& {Huntress}, Jr., W.~T. 1980, \apj, 239, 151

\bibitem[{{Regan} {et~al.}(1999){Regan}, {Sheth}, \& {Vogel}}]{regan99}
{Regan}, M.~W., {Sheth}, K., \& {Vogel}, S.~N. 1999, \apj, 526, 97

\bibitem[{{Rodriguez-Franco} {et~al.}(1998){Rodriguez-Franco},
  {Martin-Pintado}, \& {Fuente}}]{rodriguez98}
{Rodriguez-Franco}, A., {Martin-Pintado}, J., \& {Fuente}, A. 1998, \aap, 329,
  1097

\bibitem[{{Sakai} {et~al.}(2009){Sakai}, {Sakai}, {Hirota}, {Burton}, \&
  {Yamamoto}}]{sakai09}
{Sakai}, N., {Sakai}, T., {Hirota}, T., {Burton}, M., \& {Yamamoto}, S. 2009,
  \apj, 697, 769

\bibitem[{{Sakai} {et~al.}(2008){Sakai}, {Sakai}, {Hirota}, \&
  {Yamamoto}}]{sakai08}
{Sakai}, N., {Sakai}, T., {Hirota}, T., \& {Yamamoto}, S. 2008, \apj, 672, 371

\bibitem[{{Sanders} {et~al.}(2003){Sanders}, {Mazzarella}, {Kim}, {Surace}, \&
  {Soifer}}]{sanders03}
{Sanders}, D.~B., {Mazzarella}, J.~M., {Kim}, D.-C., {Surace}, J.~A., \&
  {Soifer}, B.~T. 2003, \aj, 126, 1607

\bibitem[{{Schilke} {et~al.}(1992){Schilke}, {Walmsley}, {Pineau Des Forets},
  {Roueff}, {Flower}, \& {Guilloteau}}]{schilke92}
{Schilke}, P., {Walmsley}, C.~M., {Pineau Des Forets}, G., {et~al.} 1992, \aap,
  256, 595

\bibitem[{{Solomon} {et~al.}(1997){Solomon}, {Downes}, {Radford}, \&
  {Barrett}}]{solomon97}
{Solomon}, P.~M., {Downes}, D., {Radford}, S.~J.~E., \& {Barrett}, J.~W. 1997,
  \apj, 478, 144

\bibitem[{{Sorai} {et~al.}(2002){Sorai}, {Nakai}, {Kuno}, \&
  {Nishiyama}}]{sorai02}
{Sorai}, K., {Nakai}, N., {Kuno}, N., \& {Nishiyama}, K. 2002, \pasj, 54, 179

\bibitem[{{Spoon} {et~al.}(2002){Spoon}, {Keane}, {Tielens}, {Lutz},
  {Moorwood}, \& {Laurent}}]{spoon02}
{Spoon}, H.~W.~W., {Keane}, J.~V., {Tielens}, A.~G.~G.~M., {et~al.} 2002, \aap,
  385, 1022

\bibitem[{{Spoon} {et~al.}(2007){Spoon}, {Marshall}, {Houck}, {Elitzur}, {Hao},
  {Armus}, {Brandl}, \& {Charmandaris}}]{spoon07}
{Spoon}, H.~W.~W., {Marshall}, J.~A., {Houck}, J.~R., {et~al.} 2007, \apjl,
  654, L49

\bibitem[{{Stacey} {et~al.}(1991){Stacey}, {Geis}, {Genzel}, {Lugten},
  {Poglitsch}, {Sternberg}, \& {Townes}}]{stacey91}
{Stacey}, G.~J., {Geis}, N., {Genzel}, R., {et~al.} 1991, \apj, 373, 423

\bibitem[{{Szczepanski} {et~al.}(2005){Szczepanski}, {Wang}, {Doughty}, {Cole},
  \& {Vala}}]{szczepanski05}
{Szczepanski}, J., {Wang}, H., {Doughty}, B., {Cole}, J., \& {Vala}, M. 2005,
  \apjl, 626, L69

\bibitem[{{Tielens} \& {Hollenbach}(1985)}]{tielens85}
{Tielens}, A.~G.~G.~M. \& {Hollenbach}, D. 1985, \apj, 291, 722

\bibitem[{{Wang} {et~al.}(2004){Wang}, {Henkel}, {Chin}, {Whiteoak}, {Hunt
  Cunningham}, {Mauersberger}, \& {Muders}}]{wang04}
{Wang}, M., {Henkel}, C., {Chin}, Y.-N., {et~al.} 2004, \aap, 422, 883

\end{thebibliography}

\clearpage

\begin{appendix}
\section{Line ratios and source sizes}
\label{app:nearby}
We have calculated the line ratios between the observed HC$_3$N, HCN, and HNC lines, and the used method will be demonstrated in this appendix. The method accounts for beam size/source size effects, using the following formula for a line ratio between the spectral lines A and B:

\begin{equation}
\indent \frac{I(\mathrm{A})}{I(\mathrm{B})} = \frac{{\theta_{\mathrm{mb}}}_{\mathrm{A}}^2 + \theta_s^2}{{\theta_{\mathrm{mb}}}_{\mathrm{B}}^2 + \theta_{\mathrm{s}}^2} \frac{{\eta_{\mathrm{mb}}}_{\mathrm{B}}}{{\eta_{\mathrm{mb}}}_{\mathrm{A}}} \frac{\int {T_{\mathrm{A}}^*}_{\mathrm{A}} dv}{\int {T_{\mathrm{A}}^*}_{\mathrm{B}} dv},
\label{eq:ratio}
\end{equation}

\noindent
where $\theta_{\mathrm{mb}}$ is the main beam half-power beam width (or main beam size) of the telescope, $\eta_{\mathrm{mb}}$ is the main beam efficiency of the telescope, $\theta_{\mathrm{s}}$ is the source size of the observed object, and $\int T_{\mathrm{A}}^* dv$ is the integrated intensity of the signal. The temperature can also be given in $T_{\mathrm{mb}}$ scale -- in this case, the main beam efficiency corresponding to that observation should be omitted since $T_{\mathrm{mb}} = \frac{T_{\mathrm{A}}^*}{\eta_{\mathrm{mb}}}$.

Some of the galaxies have very large angular distributions. The measured intensities might in these few cases not represent a global value for molecular gas in the galaxy, but rather a value for a certain (central) region of the galaxy. This will be discussed in Section~\ref{app:nearby2} If the different observations for the same galaxy are made in different parts of the galaxies the line ratios might be misleading. The exact positions of observation are given in most of the references, and these have been compared. The largest difference in position between two observations in the same galaxy is for \object{IC 342}, which has a 3$''$ difference between the HCN observation in \citet{sorai02} and the HNC and HC$_3$N observations in \citet{henkel88}. Compared to the sizes of HNC and HC$_3$N distribution \citep{meier05} as well as the used beam sizes (19$''$ and 25$''$), this offset is although rather small. For all other galaxies, the position difference is at most 1$''$.

Another issue is that two molecules compared in a line ratio might not have the same spatial distributions, and too narrow a beam might exclude more flux from one molecule than from the other. The problem gets even greater if the two molecular lines are observed with two different beam sizes. The problem mainly affect the most nearby galaxies in the sample, whose source sizes are comparable with the used beam sizes.

The size of the dense molecular region is also important when converting from measured temperature to brightness temperature, and as a consequence of this also when computing the line ratios between two transitions in a galaxy when different telescopes have been used. When available, the source size has been estimated from tables or maps showing the HCN source size, thus assuming that the HCN source size is similar to the source sizes of the other molecules used in the line ratio calculations (HNC and HC$_3$N) as all these lines are expected to be present only in the dense molecular regions \citep{meier05}. When an HCN source size has not been found, the source size of the CO emission has been used instead. Preference has then been given to the higher CO transitions, as these are better tracers of dense gas than CO~1\nobrkhyph0, which traces much thinner molecular gas, giving too high a value on the source size. The source sizes used in this work are shown in Table~\ref{tab:galaxies}, and unless no transition is stated, HCN~1\nobrkhyph0 should be assumed.

For some of the galaxies no reliable value on the source size (neither HCN nor CO) has been found. This might still not be a problem when calculating the line ratios. In Section~\ref{app:unknownsize} it will be shown that the error when assuming a point source is less than 5~$\%$ for objects more distant than 45~Mpc, and less than 10~$\%$ for objects more distant than 30~Mpc. These errors are calculated for a line ratio made with the most different beam sizes used in this work. When the beam sizes of the telescopes used for the observations are (almost) the same, the source sizes will just cancel. In any case, this error should be smaller than errors introduced by the use of so many different telescopes and instruments.

\subsection{Nearby galaxies}
\label{app:nearby2}

As mentioned above, some of the most nearby galaxies have source sizes larger than the beam size of the telescope used for the observation. When calculating molecular line ratios, this might pose a problem. Only if we expect the distributions of the two molecules to have the same shape and size, and the observations are made with the same beam sizes, we will achieve the same ratio as for a global measurement on the galaxy. In particular, we expect the HC$_3$N to be concentrated in a smaller region of the galaxy than HCN and HNC \citep{meier05}, why the calculations of these ratios depend on the beam size to be large enough to cover the whole dense molecular region of the galaxy (e.g. the whole HCN region).

If the beam size is smaller than the HCN (or HNC) emitting region, but larger than the HC$_3$N region, the HC$_3$N/HCN (or HC$_3$N/HNC) ratio will be overestimated, as all HC$_3$N will be seen, but not all HCN (or HNC). However, if the opposite would be true, the HC$_3$N/HCN and HC$_3$N/HNC instead would be underestimated. But since HCN and HNC are more abundant than HC$_3$N in all studied sources, this seems very unlikely.

Not knowing the HC$_3$N source size will also affect the line ratios from the more distant galaxies to some extent, since the source size used for the line ratio calculations is an HCN source size (in a few cases even a CO source size) also for the HC$_3$N intensities. However, assuming the proportion between the HCN and HC$_3$N source sizes to be similar for all galaxies, this will affect all line ratios in the same way, thus making all line ratios a little bit too high.

Another problem for the line ratios of the nearby galaxies is that the two different observations used to calculate a line ratio sometimes are made with different beam sizes. When comparing HCN~1\nobrkhyph0, HNC~1\nobrkhyph0, and HC$_3$N~10\nobrkhyph9 observations made with the same telescopes the difference in source size is negligible, but if a line ratio is calculated from observations from two different telescopes the two molecular intensities are observed towards regions with different sizes. For distant galaxies this is not a problem, since the whole molecular region of the galaxy is unresolved in any beam. For the more nearby galaxies, one of the molecules might be observed more or less globally in the galaxy, while the other is observed very locally in the galactic centre, giving an erroneous line ratio. The galaxies affected by this issue should be the same as those affected by the previously mentioned beam size issue.

\subsection{Unknown source sizes}
\label{app:unknownsize}

The source sizes used in this study are found in Table~\ref{tab:galaxies}. However, for some sources no reliable value for the source size has been found. We will here discuss why this is not always a problem, and estimate sizes of the errors inflicted from not knowing the source size.

As can be seen above, the line ratios depend on the source and beam sizes with the factor $\frac{{\theta_{\mathrm{mb}}}_{\mathrm{A}}^2 + \theta_{\mathrm{s}}^2}{{\theta_{\mathrm{mb}}}_{\mathrm{A}}^2 + \theta_{\mathrm{s}}^2}$. The source size $\theta_{\mathrm{s}}$ can be ignored if both beam sizes ${\theta_{\mathrm{mb}}}_{\mathrm{A}}, {\theta_{\mathrm{mb}}}_{\mathrm{B}} \gg \theta_{\mathrm{s}}$. If the beam sizes ${\theta_{\mathrm{mb}}}_{\mathrm{A}} \approx {\theta_{\mathrm{mb}}}_{\mathrm{B}}$, the source size will also cancel. However, if $\theta_{\mathrm{s}}$ is comparable to the beam sizes, and ${\theta_{\mathrm{mb}}}_{\mathrm{A}}$ and ${\theta_{\mathrm{mb}}}_{\mathrm{B}}$ are non-similar, the source size becomes an important factor.

The error when assuming the source to be point-like, e.g. setting $\theta_{\mathrm{s}} = 0$, will be

\begin{equation}
\indent \frac{{\theta_{\mathrm{mb}}}_{\mathrm{A}}^2 / {\theta_{\mathrm{mb}}}_{\mathrm{B}}^2}{\frac{{\theta_{\mathrm{mb}}}_{\mathrm{A}}^2 + \theta_{\mathrm{s}}^2}{{\theta_{\mathrm{mb}}}_{\mathrm{B}}^2 + \theta_{\mathrm{s}}^2}}.
\label{eq:error}
\end{equation}

\noindent
For the sources where an HCN source size has been used, the corresponding source diameter has never exceeded 1.6~kpc (\object{NGC 2146}). For the CO source sizes, the largest is found in \object{NGC 5135}, with a source diameter of 4.1~kpc in CO~1\nobrkhyph0, but the HCN source sizes should be smaller than this.

Assuming no larger HCN source diameter than 1.6~kpc, and the largest and smallest beam sizes at 90~GHz (SEST with 57$''$ and IRAM with 28$''$), the error will be less than 5~$\%$ for distances greater than 45~Mpc, and less than 10~$\%$ for distances greater than 30~Mpc. Thus, even for objects closer than 30~Mpc, the error caused by the point-like approximation ($\theta_{\mathrm{s}} = 0$) is notable only if the $\theta_{\mathrm{mb}}$ of the two observations used to calculate the ratio have notably different sizes (at least a 30~$\%$ difference in beam sizes is needed to produce an error of 10~$\%$).

\end{appendix}

\end{document}